\documentclass[aps,prb,amsmath,amssymb,showpacs,eqsecnum]{revtex4}
\usepackage[dvipdfm]{graphicx}
\usepackage{fancybox}
\usepackage{amsmath,amssymb,bm,amsthm}
\def\beq{\begin{equation}}
\def\eeq{\end{equation}}

\def\<{\langle}
\def\>{\rangle}

\begin{document}


\title{Equilibrium, Metastability, and Hysteresis in a Model Spin-crossover Material 
with Nearest-neighbor Antiferromagnetic-like and Long-range Ferromagnetic-like Interactions}

\author{Per Arne Rikvold,$^1$ Gregory Brown,$^2$ Seiji Miyashita,$^{3,4}$ Conor Omand,$^{3,5}$ 
Masamichi Nishino$^6$}
\affiliation{$^1$ Department of Physics, Florida State University, 
Tallahassee, FL 32306-4350,USA\\
$^2$ Computational Science and Mathematics Division, Oak Ridge National Laboratory, Oak Ridge, Tennessee 37830, USA\\
$^3$ Department of Physics, Graduate School of Science,
The University of Tokyo, 7-3-1 Hongo, Bunkyo-ku, Tokyo 113-8656, Japan\\
$^4$ CREST, JST, 4-1-8 Honcho Kawaguchi, Saitama, 332-0012, Japan\\
$^5$ Department of Physics and Astronomy, University of British Columbia, Vancouver, BC, V6T 1Z4, Canada\\
$^6$ Computational Materials Science Center, National Institute for Materials Science, Tsukuba, Ibaraki 305-0047, Japan
}

\begin{abstract}
Phase diagrams and hysteresis loops were obtained by Monte Carlo simulations and a mean-field method for a 
simplified model of a spin-crossover material with a two-step transition between the high-spin and low-spin states. This 
model is a mapping onto a square-lattice $S=1/2$ Ising model with 
antiferromagnetic nearest-neighbor and ferromagnetic Husimi-Temperley (equivalent-neighbor) 
long-range interactions. 
Phase diagrams obtained by the two methods for weak and strong long-range interactions are found to be similar. 
However, for intermediate-strength long-range interactions, 
the Monte Carlo simulations show that tricritical points decompose into pairs of critical endpoints and mean-field 
critical points surrounded by horn-shaped regions of metastability. Hysteresis loops along paths 
traversing the horn regions are strongly reminiscent of thermal two-step transition loops with hysteresis, 
recently observed experimentally in several spin-crossover materials. 
We believe analogous phenomena should be observable in experiments and simulations for many systems 
that exhibit competition between local antiferromagnetic-like interactions and long-range ferromagnetic-like 
interactions caused by elastic distortions. 
\end{abstract}

\pacs{75.30.Wx,64.60.My,64.60.Kw,75.60.-d}


\maketitle


\section{Introduction}
\label{sec:I}

In many materials, local elastic interactions induce effective long-range 
interactions via the macroscopic strain field.\cite{TEOD82,ZHUX05} 
In addition to elastic crystals, physical realizations of long-range interactions include systems as diverse as 
earthquake faults \cite{KLEI07} and long-chain polymers.\cite{BIND84} 
Phase transitions in such systems belong to the {\it mean-field universality class\/}, which 
has some unusual properties. In particular, critical clusters can be strongly suppressed compared to 
transitions caused by purely short-range interactions. This effect should be 
experimentally observable as an absence of critical opalescence.\cite{MIYA08,NAKA11,NAKA12}

A particular class of systems that exemplify these interesting properties are {\it spin-crossover\/} (SC) 
materials.\cite{MIYA08,NAKA11,NAKA12,ENAC05,NISH07,KONI08,BOUS11,HALC13} 
These are molecular crystals in which the individual organic molecules contain 
transition metal ions, such as Fe(II), Fe(III), or Co(II), that can exist
in two different spin states: a low-spin ground state (LS) and
a high-spin excited state (HS). Molecules in the HS state have 
higher effective degeneracy and larger volume than those in the LS state. 
Due to the higher degeneracy of the excited HS state, crystals of such
molecules can be brought into a majority HS
state by increasing temperature, changing pressure or magnetic field,
or by exposure to light.\cite{ENAC05,GUTL94,CHON10,OHKO11,ASAH12,CHAK14}
If the intermolecular interactions are sufficiently strong, this change
of state can become a discontinuous phase transition
such that the HS phase becomes metastable and hysteresis occurs.\cite{GUTL94,MIYA05}
In the case of optical excitation into the metastable phase, this effect is known as light-induced
excited spin-state trapping (LIESST).\cite{MIYA09,GUTL94}
The different magnetic
and optical properties of the two phases make such cooperative
SC materials promising candidates for applications such as switches, displays, memory devices, sensors, 
and activators.\cite{OHKO11,CHAK14,KAHN98,LINA12} 

Various experimental results over the last decade have led to the suggestion that
the dominant interaction mechanism in SC materials is elastic and therefore effectively long-range,
due to the larger size of the molecule in its HS state.\cite{NISH07,SPIE04}
Such systems can be modeled by a pseudo-spin  Hamiltonian of the form 
\begin{equation}
{\mathcal H}= -J \sum_{\langle i,j \rangle} s_i s_j 
- \frac{1}{2}  (k_{\rm B} T \ln g - D) \sum_i  s_i
+ {\mathcal H}_{\rm LR} \;.
\label{eq:ham1}
\end{equation}
The first two terms constitute the Wajnflasz-Pick Ising-like model,\cite{WAJN71} 
in which the pseudo-spin variables $s_i$
denote the two spin states ($-1$ for LS and $+1$ for HS), and $J$ is a nearest-neighbor interaction. 
The effective field term, 
\begin{equation}
H =  (k_{\rm B} T \ln g - D)/2 \;,
\label{eq:heff}
\end{equation}
contains $D > 0$, which is the energy difference between the HS and LS states, 
$g$, which is the ratio between the HS and LS degeneracies, and $k_{\rm B} T$, the absolute temperature in energy 
units. The long-range interactions are represented by $ {\mathcal H}_{\rm LR}$. 
The usual order parameter for SC materials is the proportion of HS molecules, $n_{\rm HS}$, which is related to the 
pseudo-spin variables as  $n_{\rm HS} = \left( m + 1 \right)/2$, where $m = \sum_i s_i /N$ 
(with $N$ the total number of molecules) is the pseudo-magnetization. 

While the long-range term $ {\mathcal H}_{\rm LR}$ is most often considered as a true elastic 
interaction,\cite{MIYA08,NAKA12,NISH13} 
many aspects of the model can be obtained at much lower computational cost by 
replacing this by a Husimi-Temperley (a.k.a.\ equivalent-neighbor) effective Hamiltonian.\cite{MORI10,NAKA11}
The latter is the approach we follow in the present paper. 

In previous work we have considered models of SC materials that show a direct, or one-step, transition between the 
LS and HS phases. In these cases, the short-range interactions favor configurations in which nearest-neighbor molecules 
are in the same state (LS-LS or HS-HS). This corresponds to a positive short-range interaction  constant $J$ in 
Eq.~(\ref{eq:ham1}). 
In the pseudo-spin language often used in the SC literature, this case is called ferromagnetic-like, 
or simply ferromagnetic. 
We emphasize that 
this is only an analogy and does {\em not\/} imply a magnetic character of the interactions. 
In the remainder of this paper, we will use the simplified terms, ferromagnetic and antiferromagnetic, 
for interactions that favor uniform and checkerboard spin-state arrangements, respectively. 

When both the short-range and long-range interactions 
are ferromagnetic, any nonzero long-range interaction has the effect of changing the 
universality class of the critical point of the LS/HS phase transition caused by the short-range interactions 
from the Ising to the mean-field universality class.\cite{NAKA11}  
As a result, critical clusters are suppressed, and the system develops true metastable phases limited 
by sharp spinodal lines in the phase diagram. 

There also exist SC materials, in which the transition between the LS and HS phases proceeds as a 
two-step transition via an intermediate 
phase.\cite{KOPP82,ZELE85,PETR87,BOUS92,JAKO92,REAL92,BOIN94,BOLV96,CHER04,BONN08,PILL12,BURO10,LIN12,KLEI14,HARD15} 
For some materials, such as Fe(II)[2-picolylamine]$_3$Cl$_2 \cdot$Ethanol,\cite{KOPP82}  
it has been shown by x-ray diffraction that spontaneous symmetry breaking induces an intermediate phase, 
characterized by long-range order on two interpenetrating sublattices with nearest-neighbor 
molecules in different states (HS-LS).\cite{CHER03,HUBY04}
This situation can be modeled by the Ising-like model of Eq.~(\ref{eq:ham1}) with 
antiferromagnetic nearest-neighbor interactions ($J <0$). 
Various mean-field approximations to this 
model have been considered, both with\cite{ZELE85,BOUS92,CHER04} and without\cite{BOLV96} 
a long-range ferromagnetic term. 

In a recent work, Nishino and Miyashita studied such a model using an elastic long-range interaction.\cite{NISH13} 
They found that, 
for weak applied field, the antiferromagnetic phase transition remains in the Ising universality class. At low 
temperatures they observed tricritical points separating lines of second-order and first-order phase transitions. 
However, much of their high-temperature analysis replaced the 
short-range interactions by a two-sublattice mean-field model, which neglects the effects of local fluctuations. 
The aim of the present paper is to investigate the effects of such fluctuations in an Ising-like model with 
nearest-neighbor antiferromagnetic interactions and a long-range ferromagnetic interaction of the Husimi-Temperley 
form. This enables us to obtain excellent long-time statistics for systems with over one million individual pseudo-spins, 
and for several different values of the long-range interaction strength. 
Some preliminary results were included in Ref.~\onlinecite{BROW14}. 

The organization of the rest of the paper is as follows. 
In Sec.~\ref{sec:H} we present the Ising Hamiltonian with antiferromagnetic nearest-neighbor interactions and 
a ferromagnetic 
Husimi-Temperley type long-range interaction of adjustable strength. Here we also obtain ground-state diagrams 
(zero-temperature phase diagrams) including phase coexistence points and spinodal points for different strengths of 
the long-range interaction term. 
In  Sec.~\ref{sec:MFA} we present results for a mean-field approximation to this model, in which the nearest-neighbor 
antiferromagnetic interactions are replaced by a two-sublattice mean-field approximation. Two different strengths of the 
long-range ferromagnetic interactions, which yield qualitatively different phase diagrams, are considered. 
In Sec.~\ref{sec:MC} we return to our original, nearest-neighbor antiferromagnetic interactions, investigating the 
resulting phase diagrams by Metropolis importance-sampling Monte Carlo (MC) simulations. We find that, in a certain 
range of the long-range interaction strength, the resulting phase diagrams are qualitatively different from those 
obtained by the mean-field approximations. 
In particular, tricritical points predicted by the mean-field approximations are found by MC to decompose into pairs of 
critical endpoints and mean-field critical points surrounded by horn-shaped regions of metastability. 
Here we also present hysteresis curves that are particularly relevant to 
the system's interpretation as a model of two-step transitions in SC materials. 
Our conclusions and suggestions for future work are presented in Sec.~\ref{sec:CONC}.

\section{Hamiltonian and Ground-state Analysis}
\label{sec:H}

The square-lattice Ising antiferromagnet with weak, long-range (Husimi-Temperley) ferromagnetic interactions 
is defined by the Hamiltonian
\beq
{\mathcal H}= -J \sum_{\langle i,j \rangle} s_i s_j - N \left( H + \frac{A}{2} m \right) m
\label{eq:ham}
\eeq
with $J < 0$ and $A > 0$. 
Here, $H$ is the applied field, $s_i=\pm 1$, and $m = N^{-1} \sum_i s_i$. For a square lattice of side $L$, 
$N = L^2$.  The strength of the 
long-range interaction is $A$. It is defined such that the critical temperature of the pure long-range 
ferromagnet ($J=0$) equals $A/k_{\rm B}$, where $k_{\rm B}$ is Boltzmann's constant. 
For convenience we will hereafter use the dimensionless variables, $h = H/|J|$, $a = A/|J|$, and $t = k_{\rm B} T/|J|$.  

The model's equilibrium and metastable phases at zero temperature are found by a simple 
ground-state analysis. The per-site energies of the fully ordered antiferromagnetic (AFM) 
and field-induced ferromagnetic (FM) phases are given by 
$E_{\rm AFM}/(N|J|) = - 2$, $E_+/(N|J|) = 2 - a/2 - h$, and $E_-/(N|J|) = 2 - a/2 +h$, respectively. 
Equating the AFM and FM energies yields the zero-temperature transition values of $h$ as
\beq
h_{\rm AFM/+} = 4 - a / 2
\label{eq:Hplus}
\eeq
and
\beq
h_{\rm AFM/-} = -4 + a / 2 \;.
\label{eq:Hminus}
\eeq
For $a > 8$, the AFM ground state disappears, and the system in equilibrium 
has a direct transition between the $m=-1$ and $m=+1$ FM ground states at $h=0$. 
(For $a=8$, the AFM and both FM ground states are degenerate at $h=0$.)




The limits of local stability of metastable phases at $t=0$ (zero-temperature spinodal fields) are the field values 
at which the energy change due to a flip of a single spin in the metastable phase becomes negative.\cite{NEVE91} 
For metastability, a positive energy change is required.
The spinodal fields are determined as follows. 

\subsection{$a < 8$}

Decreasing $h$, attempting to nucleate 
a transition from metastable $m=+1$ to the AFM or $m = -1$ ground state by a single spin flip,
\beq
\begin{array}{ccccccccc}
+ & + & + & & & & + & + & + \\
+ & + & + & & \rightarrow & & + & - & +\\
+ & + & + & & & & + & + & + \\ \nonumber
\end{array}
\;,
\eeq
the energy change is
\beq
\Delta E/|J| = -8 +2h +2 a (1-1/N) > 0 \Rightarrow h > 4 - a (1-1/N) \;.
\label{eq:metaFA}
\eeq
Thus, (in the limit $N \rightarrow \infty$) the $m=+1$ phase is metastable  
for $4 - a < h < 4 - a/2$.
By symmetry, the $m=-1$ phase is  metastable for  $- 4 + a/2 < h < -4 + a$.

Increasing $h$, attempting to nucleate a transition from metastable AFM to the $m=+1$ ground state by a single spin flip,
\beq
\begin{array}{ccccccccc}
- & + & - & & & & - & + & - \\
+ & - & + & & \rightarrow & & + & + & +\\
- & + & - & & & & - & + & - \\ \nonumber
\end{array}
\;,
\eeq
the energy change is
\beq
\Delta E/|J| = 8 -2h -2 a /N > 0 \Rightarrow h < 4 - a/N \;.
\label{eq:metaAF}
\eeq
Thus, (in the limit $N \rightarrow \infty$) 
the AFM phase is metastable against decay to $m=+1$ for $4 - a/2 < h < 4$.
By symmetry, the AFM phase is metastable against decay to $m=-1$ for  $-4 < h < -4  + a/2$.
Four mean-field sharp spinodal lines extend from these zero-temperature spinodal points toward higher $t$. 
The zero-temperature limits of the phase diagrams shown in 
Figs.~\ref{fig:MFA_A70}(a), \ref{fig:PhD_A40}(a), and \ref{fig:PhD_A70}(a) 
illustrate the positions of coexistence and spinodal points for values of $a < 8$. 

\subsection{$a > 8$}

The only ground states are $m=+1$ for $h \ge 0$ and $m=-1$ for $h \le 0$. Increasing $h$, attempting to nucleate 
a transition from metastable $m=-1$ to the $m=+1$ ground state by a single spin flip,
\beq
\begin{array}{ccccccccc}
- & - & - & & & & - & - & - \\
- & - & - & & \rightarrow & & - & + & -\\
- & - & - & & & & - & - & - \\ \nonumber
\end{array}
\;,
\eeq
the energy change is 
\beq
\Delta E/|J| = -8 -2h +2a(1- 1 /N) > 0 \Rightarrow h <a (1-1/N)-4 \;.
\label{eq:metaAF8}
\eeq
Thus,  (in the limit $N \rightarrow \infty$) $m=-1$ is metastable for $0 < h < a -4$. By symmetry, 
$m=+1$ is metastable for $-a +4 < h < 0$. 
Using Eq.~(\ref{eq:metaAF}) and symmetry, we find that the AFM state, while never a ground state, is metastable for 
$-4 < h < 4$. 
The zero-temperature limits of the phase diagrams shown in 
Figs.~\ref{fig:MFA_A100}(a) and \ref{fig:PhD_A100}(a)
illustrate the positions of coexistence and spinodal points for $a > 8$.

\section{Mean-field Approximation}
\label{sec:MFA}

In order to obtain an approximate picture of the behavior of the model at finite $t$, we employ a simple, 
two-sublattice Bragg-Williams mean-field approximation \cite{ZELE85,BOUS92,BOLV96,BRAG34,BRAG35} with sublattice magnetizations 
$m_{\rm A} = 2N^{-1} \sum_{i \in {\rm A}} s_i$ on sublattice A ($m_{\rm A}  \in [-1,+1]$) 
and analogously for $m_{\rm B}$ on sublattice B. 
The magnetization and staggered magnetization are given by 
$m = (m_{\rm A} + m_{\rm B})/2$ and $m_{\rm Stag} = (m_{\rm A} - m_{\rm B})/2$, respectively. 
The approximation is defined by the Hamiltonian
\begin{eqnarray}
\frac{{\mathcal H}_{\rm MFA}}{N|J|} &=& \frac{4}{2} m_{\rm A} m_{\rm B} 
- \frac{a}{2} m^2 - hm \nonumber \\
 &=& \frac{1}{2} \left( 4 - \frac{a}{2} \right) m_{\rm A} m_{\rm B} 
- \frac{a}{8} (m_{\rm A}^2 +  m_{\rm B}^2 ) 
- \frac{h}{2} (m_{\rm A} +  m_{\rm B}) \;.
\label{eq:HMF}
\end{eqnarray}
Note that the effective interaction between the two sublattice magnetizations includes the long-range interaction strength $a$ and changes sign from AFM for weak $a$ to FM for strong $a$ at $a=8$. For $a=8$, the mean-field approximation describes two independent, ferromagnetic sublattices. 

The system entropy in the mean-field approximation is the sum of the two sublattice entropies,
\beq
S_{\rm MFA} = - \frac{N}{2} \sum_{{\rm X}={\rm A}}^{\rm B}
\left( 
\frac{1+m_{\rm X}}{2} \ln \frac{1+m_{\rm X}}{2} 
+ \frac{1-m_{\rm X}}{2} \ln \frac{1-m_{\rm X}}{2}  
\right) \;,
\label{eq:SMF}
\eeq
and the resulting free energy is \cite{BRAG34,BRAG35,OSOR83}
\beq
F_{\rm MFA} = {\mathcal H}_{\rm MFA} - k_{\rm B} TS_{\rm MFA}  \;.
\label{eq:FMF}
\eeq
The coupled self-consistency equations, $\partial F_{\rm MFA} / \partial m_{\rm A} = 0$ and 
 $\partial F_{\rm MFA} / \partial m_{\rm B} = 0$, become
\beq
m_{\rm A} = \tanh \left[ \frac{\left(\frac{a}{2} - 4 \right) m_{\rm B} + \frac{a}{2} m_{\rm A} + h }{t} \right]
\label{eq:mAeq}
\eeq
and equivalent for $m_{\rm B}$ with A and B interchanged on the right-hand side. 

For $a < 8$ and $h=0$, 
the global free-energy minima lie along the AFM axis ($m_{\rm B} = -m_{\rm A}$) in the order-parameter plane, 
and the self-consistency equation for the staggered magnetization becomes 
\beq
m_{\rm Stag} = \tanh \frac{4 m_{\rm Stag}}{t} 
\label{eq:mstag}
\eeq
with the N{\'e}el temperature $t_{\rm N} =4$, independent of $a$. 
 For $a > 8$ and $h=0$, 
the global free-energy minima lie along the FM axis ($m_{\rm B} = m_{\rm A}$), 
and the self-consistency equation for the magnetization becomes 
\beq
m = \tanh \frac{(a-4) m}{t} 
\label{eq:m}
\eeq
with the $a$-dependent Curie-Weiss critical temperature $t_{\rm C} =(a-4)$. 
For non-zero applied fields, free-energy minima and saddle points in the 
$(m_{\rm A},m_{\rm B})$ plane were obtained numerically using Mathematica. 
Resulting phase diagrams for $a=7$ and $a=10$ are shown in Figs.~\ref{fig:MFA_A70} and \ref{fig:MFA_A100}, respectively.

\subsection{$a = 7$}
\begin{figure}[ht]
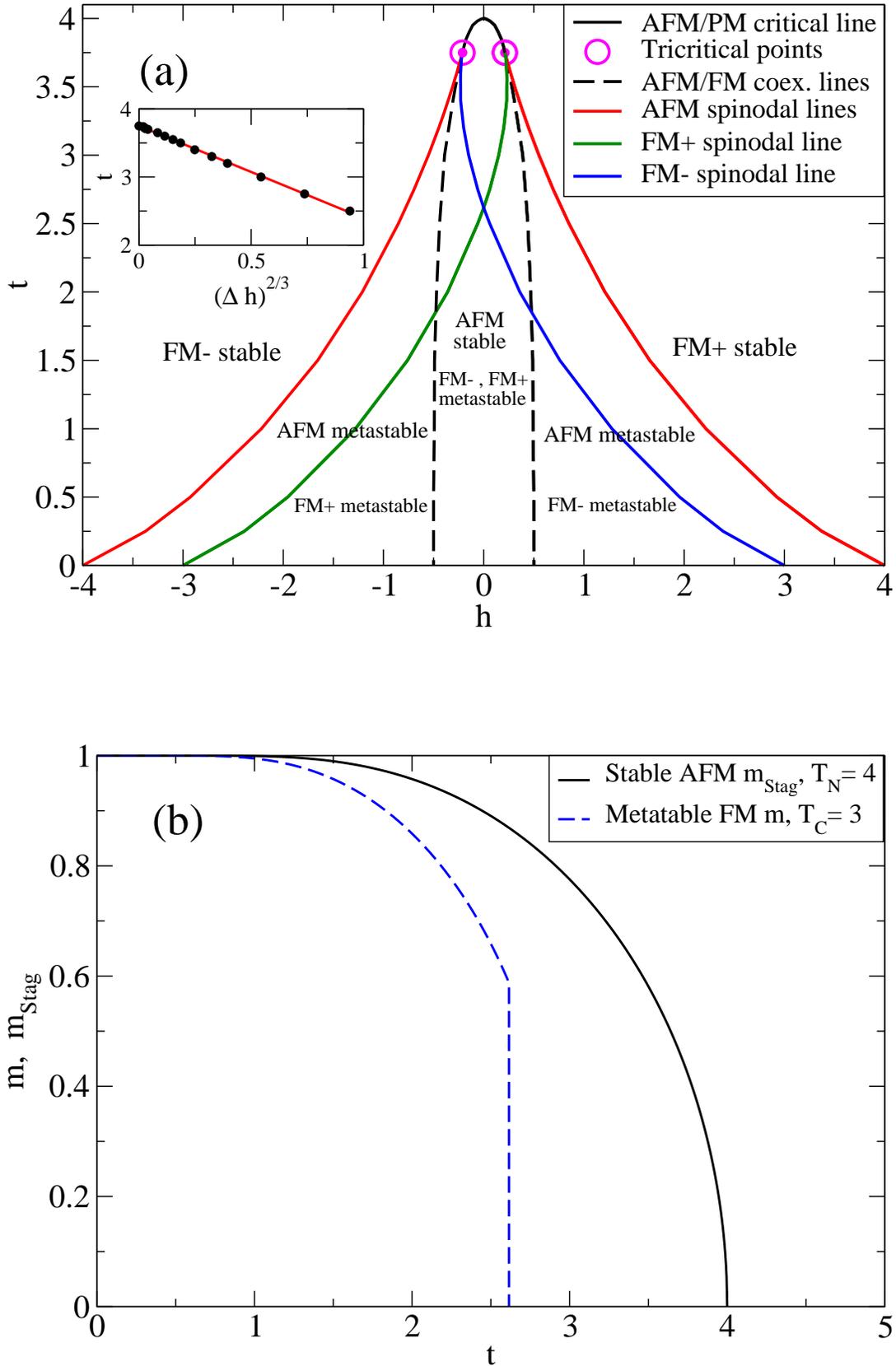

\begin{center}
\vspace*{-0.8truecm}
\includegraphics[angle=0,width=.8\textwidth]{Mean-Field_PhaseDiagram_A70.eps} \\
\vspace*{1.8truecm}
\includegraphics[angle=0,width=.8\textwidth]{H0_PhaseDiag_A70_paper.eps} 
\end{center}
\vspace*{-0.3truecm}
\caption[]{
\baselineskip=0.15truecm
(Color online) 
Mean-field phase diagrams for $a=7 < 8$. 
(a)  In the $h,t$ plane, showing a line of equilibrium critical points terminated by two tricritical points, equilibrium coexistence lines, and sharp spinodals. The FM+ phase is stable everywhere on the positive-$h$ side of the right-hand coexistence 
line, and analogously for FM$-$ on the negative-$h$ side. 
The inset demonstrates the $(t_c - t)^{3/2}$ behavior of the spinodal fields, as given by 
Eq.~(\protect\ref{eq:hspin}). Here, $\Delta h$ is the difference between the FM and corresponding AFM spinodals. 
 (b) Showing the stable AFM order parameter $m_{\rm Stag}$ and the metastable FM order parameter $m$ vs $t$ 
for $h=0$. 
The latter terminates at the spinodal temperature $t_{\rm sFM} \approx 2.615$, corresponding to the crossing of the 
two FM spinodal field curves in part (a). 
See further discussion in the text. 
}
\label{fig:MFA_A70}
\end{figure}

\begin{figure}[ht]
\begin{center}
\vspace*{-0.9truecm}
\includegraphics[angle=0,width=.5\textwidth]{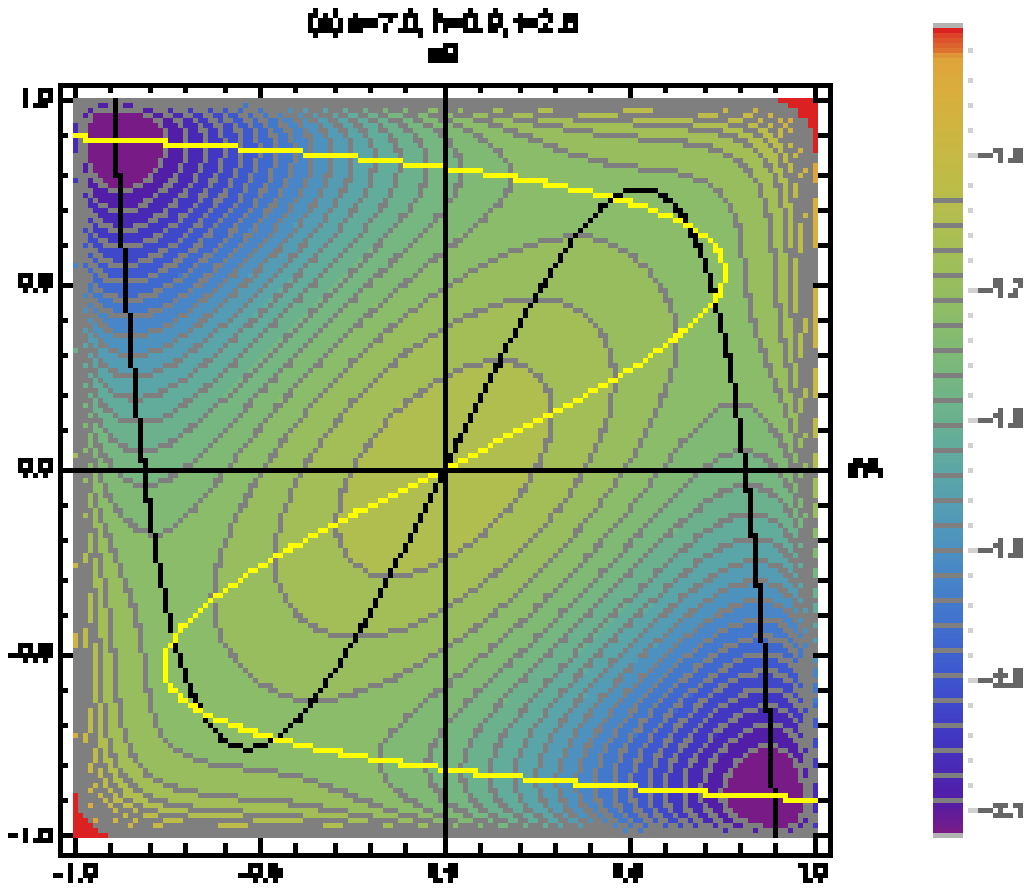} 
\includegraphics[angle=0,width=.5\textwidth]{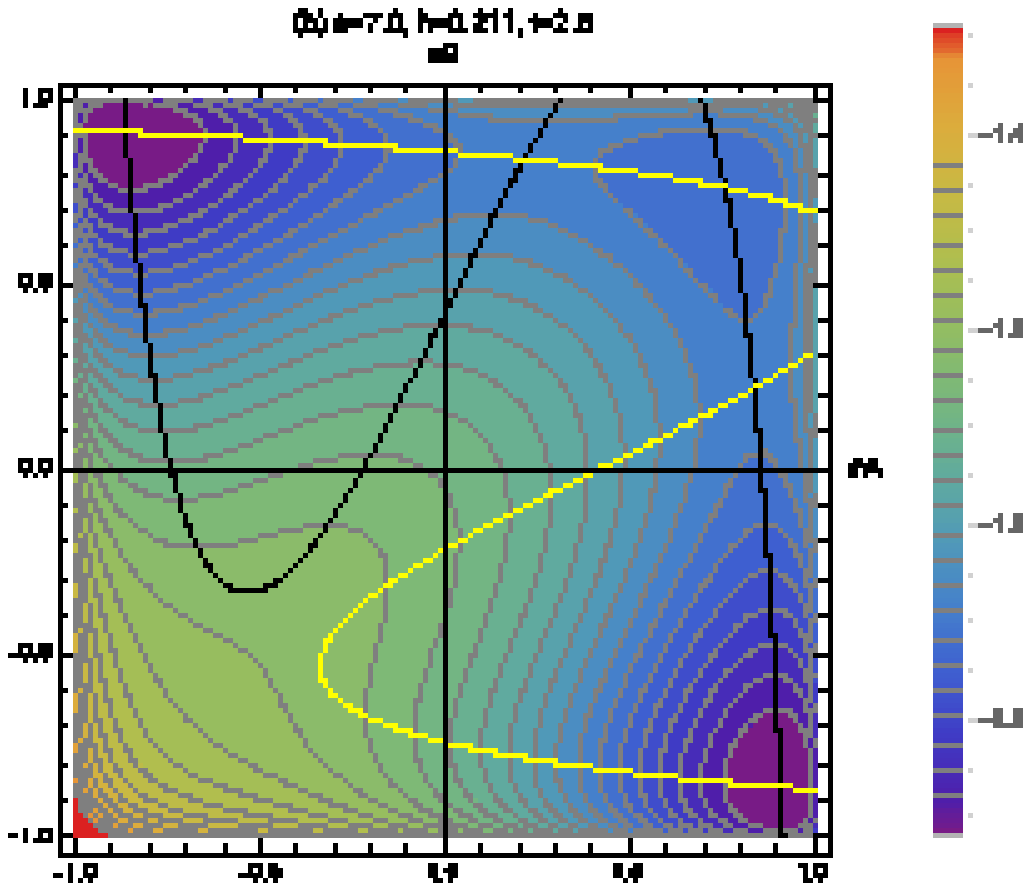} 
\includegraphics[angle=0,width=.5\textwidth]{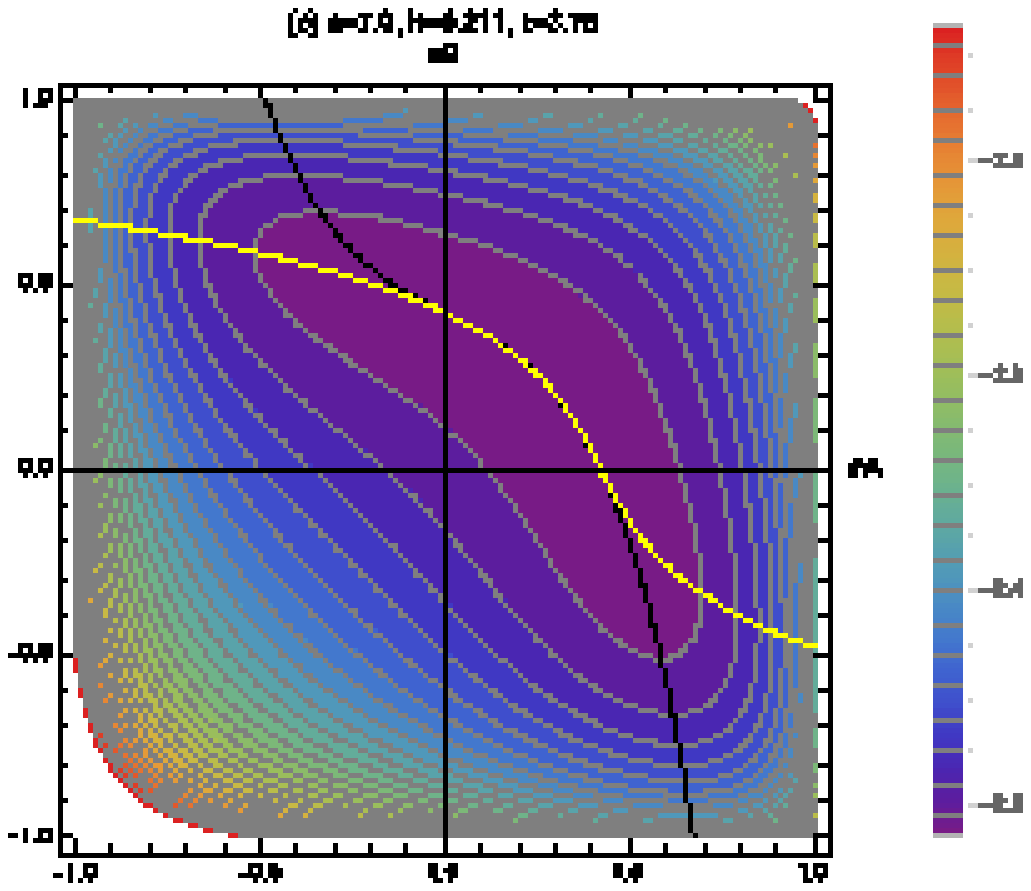}
\end{center}
\vspace*{-0.9truecm}
\caption[]{
\baselineskip=0.15truecm
(Color online) 
Contour plots in the $m_{\rm A},m_{\rm B}$ plane of the mean-field free energy for $a=7$. 
The black and light gray (yellow online) curves represent the solutions of 
$\partial F_{\rm MFA} / \partial m_{\rm A} = 0$ and $\partial F_{\rm MFA} / \partial m_{\rm B} = 0$, 
respectively. Crossings of the curves correspond to extrema and saddle points of the free-energy surfaces. 
(a)
$t=2.5$ and $h=0$. Global minima in the second and fourth quadrant represent the degenerate AFM stable phases. 
Local minima in the first and third quadrants represent the degenerate FM metastable phases. 
(b)
$t=2.5$ and $h=0.211$.  Global minima in the second and fourth quadrants represent the degenerate 
AFM stable phases. The local minimum in the first quadrant represents the metastable FM+ phase. 
(c)
$t=3.75$ and $h=0.211$, corresponding to the tricritical point where the two AFM and the FM+ phases are 
indistinguishable. The shallow, global minimum lies in the first quadrant along the FM ($m_{\rm B} = m_{\rm A}$) axis.  
}
\label{fig:MFFE}
\end{figure}

Typical mean-field phase diagrams for the model with $a<8$ are shown in  Fig.~\ref{fig:MFA_A70},
here using $a=7$. In Fig.~\ref{fig:MFA_A70}(a), the line of 
N{\'e}el critical points, marking the continuous phase transition between the stable AFM phase and the 
high-temperature disordered phase, extends from $t_{\rm N} = 4$ to two tricritical points at a lower temperature, 
which decreases with decreasing $a$.
(For $a=7$, the tricritical points are found at $t_3 \approx 3.75$ and $h_3 \approx \pm 0.211$, 
as seen in Figs.~\ref{fig:MFA_A70}(a) and \ref{fig:MFFE}(c).) 
Lines of first-order phase transitions (AFM/FM coexistence lines) extend from the tricritical points to the 
zero-temperature coexistence points at $h = \pm (4 - a/2) = \pm 0.5$. Sharp spinodal field 
lines marking the limits of metastability for 
the AFM phase extend from the tricritical points to the zero-temperature spinodal points at $h = \pm 4$, and spinodals 
marking the limits of metastability for the FM+ and FM$-$ phases extend to their zero-temperature termination points at 
$h = 4-a =-3$ and $h = -4+a = +3$, respectively. 
The corresponding zero-field order parameters, the equilibrium AFM $m_{\rm Stag}$ and 
the metastable FM $m$, which follow Eqs.~(\ref{eq:mstag}) and (\ref{eq:m}), respectively,
are shown in  Fig.~\ref{fig:MFA_A70}(b). However, the free-energy barriers that prevent the decay of the 
metastable FM phases (which are possible at $h=0$ only for $4 < a < 8$)  
into an AFM phase vanish at a spinodal temperature $t_{\rm sFM} \approx 2.615$, marked 
by a discontinuous drop of $m$ to zero. 
This temperature corresponds to the crossing of the two FM spinodals in  Fig.~\ref{fig:MFA_A70}(a).
Above it, the FM solutions of the self-consistency equations  become saddle points. 
Samples of free-energy contour plots in the $m_{\rm A}, m_{\rm B}$ plane, based on Eq.~(\ref{eq:FMF}), 
overlaid with curves 
representing the individual solutions of the two self-consistency equations, Eq.~(\ref{eq:mAeq}), 
are shown in Fig.~\ref{fig:MFFE}. 

Mean-field models of two-step crossover 
have previously been considered by Zelentsov et al.,\cite{ZELE85} 
Bousseksou et al.,\cite{BOUS92} and  Bolvin,\cite{BOLV96}  with  
Bragg-Williams approaches similar to ours, and by Chernyshov et al.\cite{CHER04} using Landau theory. 
However, Bolvin does not include any FM interaction, and consequently the resulting mean-field phase diagrams 
are those of a pure Ising antiferromagnet without any first-order transitions (Figs.~5 and 6 of Ref.~\onlinecite{BOLV96}).
Zelentsov et al.\ and Bousseksou et al.\ 
confine the FM interactions to each sublattice separately, and so these interactions do not affect the 
effective inter-sublattice interaction as they do in our model [see Eq.~(\ref{eq:HMF})]. Neither paper contains 
explicit phase diagrams. However,  their plots of HS fraction vs temperature indicate that phase diagrams 
for the case of identical sublattices (their intra-sublattice interactions $J_A=J_B$) should be similar to ours, including 
tricritical points and first-order transitions at low temperatures. 
However, their intra-sublattice FM interactions lower the energies of the FM and AFM ground states by the same 
amount, with the result that they, in contrast to the corresponding term in our model, 
do not influence the ground-state diagram of the pseudo-spin model. 
In that respect, their approaches would correspond to a mean-field approximation to a square-lattice Ising model with 
AFM nearest-neighbor (inter-sublattice) and FM next-nearest neighbor (intra-sublattice) 
interactions.\cite{LAND72,LAND81,RIKV83}
In the Landau-theory approach of Chernyshov et al., the ferromagnetic interactions are not limited to 
the separate sublattices. For not too negative values of their temperature-like parameter $\alpha_2$, 
the phase diagram in  Fig.~5 of Ref.~\onlinecite{CHER04} is quite similar to our Fig.~\ref{fig:MFA_A70}(a). 
We believe an approach, in which the ferromagnetic interactions are not limited to the individual sublattices, 
provides a better approximation for the effects of a long-range elastic interaction. 

\subsection{$a = 10$}
\begin{figure}[ht]
\begin{center}
\vspace*{-0.8truecm}
\includegraphics[angle=0,width=.8\textwidth]{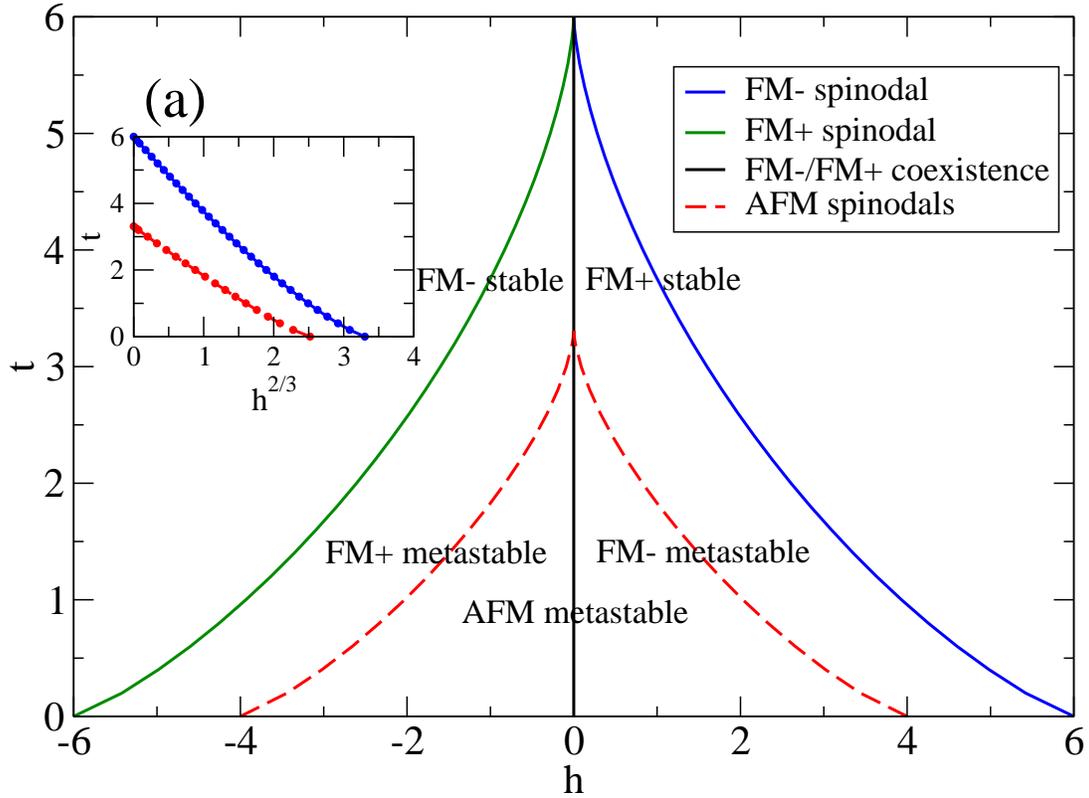} \\
\vspace*{1.8truecm}
\includegraphics[angle=0,width=.8\textwidth]{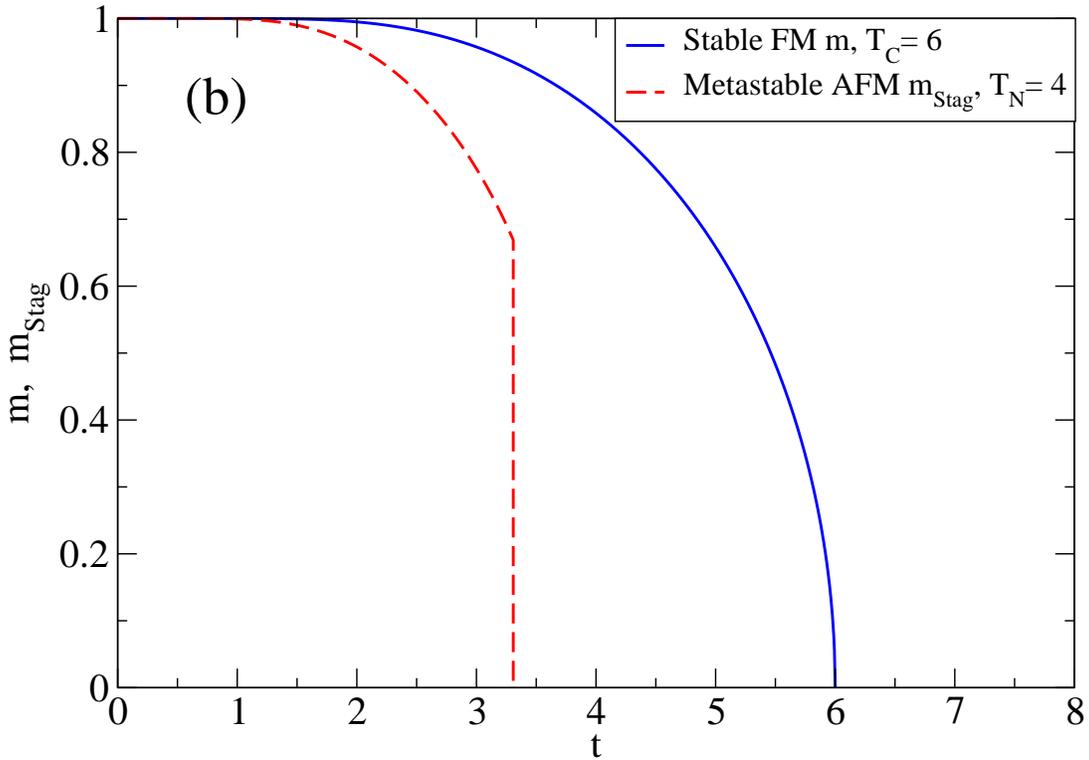} 
\end{center}
\caption[]{
\baselineskip=0.15truecm
(Color online) 
Mean-field phase diagrams for $a=10 > 8$. (a)  In the $h$,$t$ plane, showing the equilibrium coexistence line 
at $h=0$ and the spinodal field curves. 
The inset demonstrates the $(t_c - t)^{3/2}$ behavior of the spinodal fields, as given by 
Eq.~(\protect\ref{eq:hspin}). 
(b) Showing the stable FM order parameter $m$ and the metastable AFM order parameter $m_{\rm Stag}$ vs $t$ 
at $h=0$. 
The latter terminates at the spinodal temperature $t_{\rm sAFM} \approx 3.308$, corresponding to the meeting of 
the two AFM spinodal field curves in part (a).
See further discussion in the text. 
}
\label{fig:MFA_A100}
\end{figure}

Figure \ref{fig:MFA_A100} shows typical mean-field 
phase diagrams for the model with $a > 8$, here using $a=10$. There are only FM 
equilibrium phases, which coexist at $h=0$ up to the Curie temperature, $t_{\rm C} = a -4 = 6$. 
In Fig.~\ref{fig:MFA_A100}(a), sharp spinodal field lines marking the limits 
of metastability for the FM+ and FM$-$ phases extend from the critical point to their zero-temperature termination points at 
$h = 4-10 =-6$ and $h = -4+10 = +6$, respectively. Two degenerate, metastable AFM phases are possible 
at low $t$ and $h$, and the spinodals marking their limits of local stability are also shown. 
The corresponding zero-field order parameters, the equilibrium FM $m$ and the metastable AFM 
$m_{\rm Stag}$, which follow Eqs.~(\ref{eq:m}) and (\ref{eq:mstag}), respectively,
are shown in  Fig.~\ref{fig:MFA_A100}(b). However, the 
metastable AFM phases become unstable toward decay into a FM phase at a spinodal temperature $t_{s {\rm AFM}} \approx 3.308$, where the two AFM spinodals meet in  Fig.~\ref{fig:MFA_A100}(a).

In both ranges of the long-range interaction strength $a$, near the critical, tricritical, or spinodal temperature 
the spinodal fields obey the power law,\cite{NEWM80} 
\beq
|h_{\rm Spin} - h_{\rm Coex} | \sim (t_c - t)^{3/2} \;,
\label{eq:hspin}
\eeq
where $h_{\rm Spin}$ and $h_{\rm Coex}$ are the spinodal and coexistence fields, respectively, 
and $t_c$ represents the appropriate temperature where they meet. 
This is shown in the insets in Figs.~\ref{fig:MFA_A70}(a) and~\ref{fig:MFA_A100}(a).

\section{Monte Carlo simulations}
\label{sec:MC}

The standard mean-field approximation for the short-range interactions, discussed in Sec.~\ref{sec:MFA}, 
does not properly describe the microscopic 
fluctuations that are important in low-dimensional systems, especially near critical and multicritical 
points.\cite{ENDNOTE,OMAN16} 
We therefore return to the 
full model described by Eq.~(\ref{eq:ham}) to further investigate its phase diagrams and dynamics using 
importance-sampling Metropolis MC simulations. 
We consider $L \times L$ square lattices with $L = 64$, ..., 1024 and 
periodic boundary conditions. Results are extrapolated to the thermodynamic limit using known finite-size scaling 
relations. 

Critical points are located by crossings of fourth-order Binder cumulants \cite{BIND81B} 
for the antiferromagnetic order parameter $m_{\rm Stag}$, 
\beq
U_L = 1 - \frac{\langle m_{\rm Stag}^4 \rangle_L}{3 \langle m_{\rm Stag}^2 \rangle_L^2  } \;.
\label{eq:Bcum}
\eeq
(In general, the moments included in this equation are {\it central moments}, but since the model contains no staggered 
field, this is automatically satisfied for the moments of $m_{\rm Stag}$.) This method significantly reduces finite-size 
effects, and the results are further linearly extrapolated to $1/L = 0$.\cite{NAKA11} 
For isotropic interactions and periodic boundary conditions on a square lattice, as used here, 
the Ising fixed-point value of the cumulant is $U^* = 0.61...$.\cite{KAMI93,CHEN04,SELK05}

Coexistence lines represent first-order phase transitions between stable equilibrium phases, i.e., 
equality of the corresponding bulk free energies.  
Here, we locate the coexistence lines by 
starting simulations from an initial configuration consisting of two slabs, one in the 
AFM ground state and one in the FM ground state corresponding to the sign of $h$ (mixed start method), 
and searching for the field at which the final state would be either with 
approximately 50\% probability. 

The long-range, ferromagnetic interactions produce a finite  barrier in the free-energy {\it density\/}, separating the 
metastable and stable phases. As a result, the {\it total\/} free-energy barrier increases linearly 
with the system size, $N = L^2$, leading to an exponential size divergence for the lifetime of such a ``true" 
metastable phase.\cite{MORI10,TOMI76} 
(We note that this situation is radically different from the case of metastable decay in systems with only local interactions. 
In that case, the decay occurs through nucleation and growth of compact droplets of the stable phase, and 
the metastable lifetime becomes system-size {\it independent\/} in the thermodynamic limit.\cite{RIKV94A})
The sharp spinodal lines at which the free-energy barrier vanishes are 
located by starting an $L \times L$ system in the equilibrium phase and slowly scanning $h$ past the coexistence line 
(where the initial phase becomes metastable) 
until the order parameters undergo simultaneous, discontinuous jumps denoting the limit of metastability. 
The field $h_{\rm Spin}$, corresponding to the instability, 
was extrapolated to the thermodynamic limit according to the finite-size scaling relation \cite{MIYA09}
\beq
| h_{\rm Spin} - h_L | \sim L^{-4/3} \;,
\label{eq:Hspin}
\eeq
where $h_L$ is the field at which the metastable phase becomes unstable for the given value of $L$. 

\subsection{$a = 4$}
\begin{figure}[ht]
\begin{center}
\vspace*{-0.8truecm}
\includegraphics[angle=0,width=.8\textwidth]{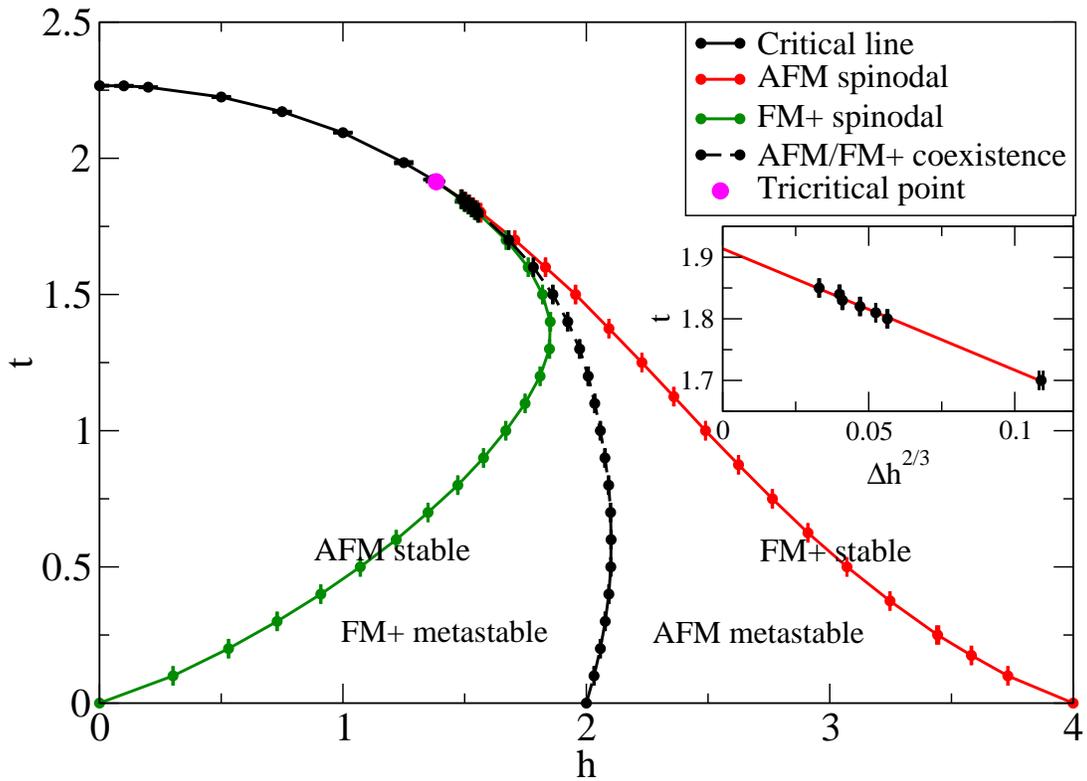}
\end{center}
\vspace*{1.5truecm}
\caption[]{
\baselineskip=0.15truecm
(Color online) 
(a) 
MC phase diagram for the full model with $a=4$. 
Except for the absence of metastable FM phases at $h=0$, the phase 
diagram resembles the mean-field phase diagram for $a=7$, shown in Fig.~\protect\ref{fig:MFA_A70}. 
However, in contrast with the mean-field model, the line of critical points 
belongs to the two-dimensional Ising universality class. At $h=0$ the critical temperature is near the exact Ising 
value, $t_c(h=0) \approx 2.269$. 
Sharp spinodal lines extend from tricritical points at $t_3 \approx 1.914$ and $h_3 \approx \pm 1.383$, separated by a 
field distance in agreement with Eq.~(\protect\ref{eq:hspin}).
The inset demonstrates this agreement near the tricritical point and was used to estimate its position. 
$\Delta h$ is defined as in Fig.~\protect\ref{fig:MFA_A70}(a). 
See further discussion in the text. 
}
\label{fig:PhD_A40}
\end{figure}

The $h,t$ phase diagram for $a=4$ is shown in Fig.~\ref{fig:PhD_A40}. 
Except for the absence of metastable FM phases at $h=0$, which is due to the lower value of $a$ used here, the phase 
diagram is topologically identical to the mean-field phase diagram for $a=7$, shown in Fig.~\ref{fig:MFA_A70}. 
However, the line of critical points belongs to the Ising universality class, as evidenced by cumulant crossing values 
$U^* \approx 0.61$. 
At $h=0$ the critical temperature is near the exact Ising value, $t_c(h=0) = 2/\ln (1 + \sqrt{2}) \approx 2.269$, unaffected by the long-range interaction.
Sharp spinodal lines extend from the tricritical point, separated by a field distance 
in agreement with Eq.~(\ref{eq:hspin}). (See inset in  Fig.~\ref{fig:PhD_A40}.)
Near the tricritical point, values of $L$ as large as 1024 were used. 
To estimate the position of the tricritical point, we extrapolated the separation between the $L$-extrapolated 
spinodal fields to zero 
according to  Eq.~(\ref{eq:hspin}) to find the tricritical temperature and from it the corresponding field values. The result is 
$t_3 \approx 1.914$ and $h_3 \approx \pm 1.383$
The coexistence line was obtained by the mixed start method with $L=512$. 
No significant differences were observed with larger $L$. 

We note that, for this relatively weak long-range interaction, our MC phase diagram shown in Fig.~\ref{fig:PhD_A40} 
is qualitatively similar to mean-field phase diagrams, both our Fig.~\ref{fig:MFA_A70} 
and Fig.~5 of Ref.~\onlinecite{CHER04}. However, for stronger long-range interactions, complex features 
that are not seen in the mean-field approximations are revealed 
by our MC simulations. These are discussed in Secs.~\ref{sec:4B} and \ref{sec:4C}  below.  

\subsection{$a = 7$}
\label{sec:4B}

\subsubsection{Phase diagram}

\begin{figure}[ht]
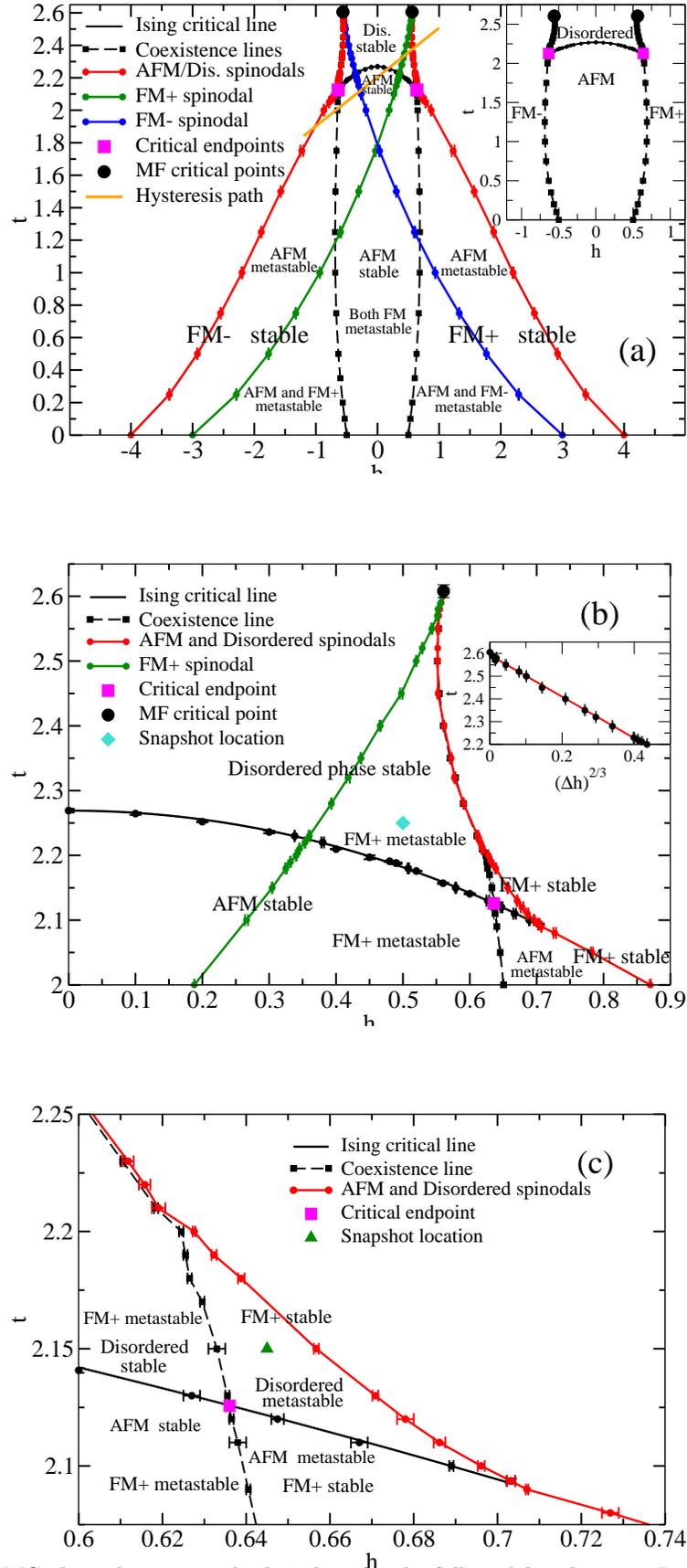

\begin{center}
\includegraphics[angle=0,width=.55\textwidth]{PhDia_A70_W_new_dataNEW.eps} \\
\vspace*{1.1truecm}
\includegraphics[angle=0,width=.55\textwidth]{PhDia_A70_W_new_data_DetailNEW.eps} \\
\vspace*{1.0truecm}
\includegraphics[angle=0,width=.55\textwidth]{PhDia_A70_W_new_data_SuperDetailNEW.eps}
\end{center}
\vspace*{-0.8truecm}
\caption[]{
\baselineskip=0.15truecm
(Color online) 
(a) 
MC phase diagram in the $h,t$ plane for the full model with $a=7$. 
In contrast with the mean-field model, the critical line is in the two-dimensional Ising universality class. 
Regions of phase stability and metastability are marked with text. 
The inset shows the phase diagram including only the stable phases. 
The diagonal line marks the path for the 
hysteresis loops in Fig.~\protect\ref{fig:Hhyst}. 
(b) Detail of the horn region of the phase diagram. 
Our estimates of the positions of the critical endpoint and the mean-field critical points are 
$t = 2.126(1)$, $h = \pm 0.636(1)$ and $t = 2.61(1)$, $h = \pm 0.561(1)$, respectively. 
At $h=0$ the critical temperature is near the exact Ising value.
The diamond marks the position of the snapshots in Fig.~\protect\ref{fig:snap}. 
The inset demonstrates that the width of the horn region, $\Delta h$, obeys Eq.~(\protect\ref{eq:hspin}). 
The straight line is a guide to the eye. 
(c) Further magnified detail of the triangular region between the coexistence line, the critical line, and the spinodal 
line for the disordered phase. Here, the stable phase is FM+ (confirmed by simulations at the point 
marked with a triangle) and the disordered phase is metastable. 
See further discussion in the text. 
}
\label{fig:PhD_A70}
\end{figure}

The MC phase diagram in the $h,t$ plane for $a=7$ is shown in Fig.~\ref{fig:PhD_A70}. 
Because of the stronger long-range interactions, metastable FM phases are possible for weak fields and low temperatures. 
However, the main difference from the case of $a=4$ (Fig.~\ref{fig:PhD_A40}) is that the tricritical points have been 
transformed into critical endpoints,\cite{WIDO77,COLL88} 
where the line of Ising critical points meets the coexistence lines at a large angle 
(light gray squares, magenta online). 
Above the temperature of the critical endpoints, the coexistence lines continue toward higher temperatures, 
each eventually terminating at a mean-field critical point (large, black circles). 
Below the critical line and between the two coexistence lines, the stable phase is AFM. On the positive 
(negative) side of the right-hand (left-hand) coexistence line, the stable phase is FM+ (FM$-$). 
Above the critical line and  between the coexistence lines, the stable phase is disordered with local fluctuations 
of AFM symmetry and a small magnetization in the direction of the applied field. 
For clarity, the inset in Fig.~\ref{fig:PhD_A70}(a) shows the phase diagram with only the stable phases and 
corresponding phase transition lines and points included. 
The critical and coexistence lines were obtained as described above, with the coexistence lines 
calculated with $L=1024$ to minimize the uncertainty. 
Our best estimates for the positions of the critical endpoints and mean-field critical points,
based on simulations up to $L = 1024$ and finite-size scaling extrapolations, 
are $t = 2.126(1)$, $h = \pm 0.636(1)$ and  $t = 2.61(1)$, $h = \pm 0.561(1)$, respectively. 
Since the full phase diagram, including metastable phase regions and spinodal lines is quite complicated, we present 
three, increasingly detailed views. 

The main part of Fig.~\ref{fig:PhD_A70}(a) shows the full range of temperatures and fields covered by the phase 
diagram, including the metastable phases and spinodal lines.
The spinodal lines marking the limits of metastability of the FM+ and FM$-$ phases cross the line of critical points at 
fields significantly weaker than those of the critical endpoints. 
Each of the FM spinodals continues on to meet the corresponding disorder spinodal at a 
mean-field critical point, forming a horn-like region of metastability. 
(As discussed below, the disorder spinodal 
and the disorder/FM coexistence line coincide within our numerical accuracy for $t \gtrsim 2.2$.) 
We located the critical point by scanning $h$ at constant $t$ across the coexistence line and monitoring the maximum of 
the susceptibility, $\chi_{\rm max}$. At the critical point $\chi_{\rm max} \sim L^{\gamma / \nu_{\rm eff}}$, with 
the mean-field exponents $\gamma = 1$ and $\nu_{\rm eff} = 2/d =1$.\cite{NAKA11,PRIV83,LUIJ97} 
Above the critical temperature, the scaling is sublinear in $L$, and below it is superlinear. 
The gray (orange online), diagonal line corresponds to the path for the hysteresis loops shown in Fig.~\ref{fig:Hhyst}. 

\begin{figure}[ht]
\begin{center}
\vspace*{-0.8truecm}
\includegraphics[angle=0,width=.4\textwidth]{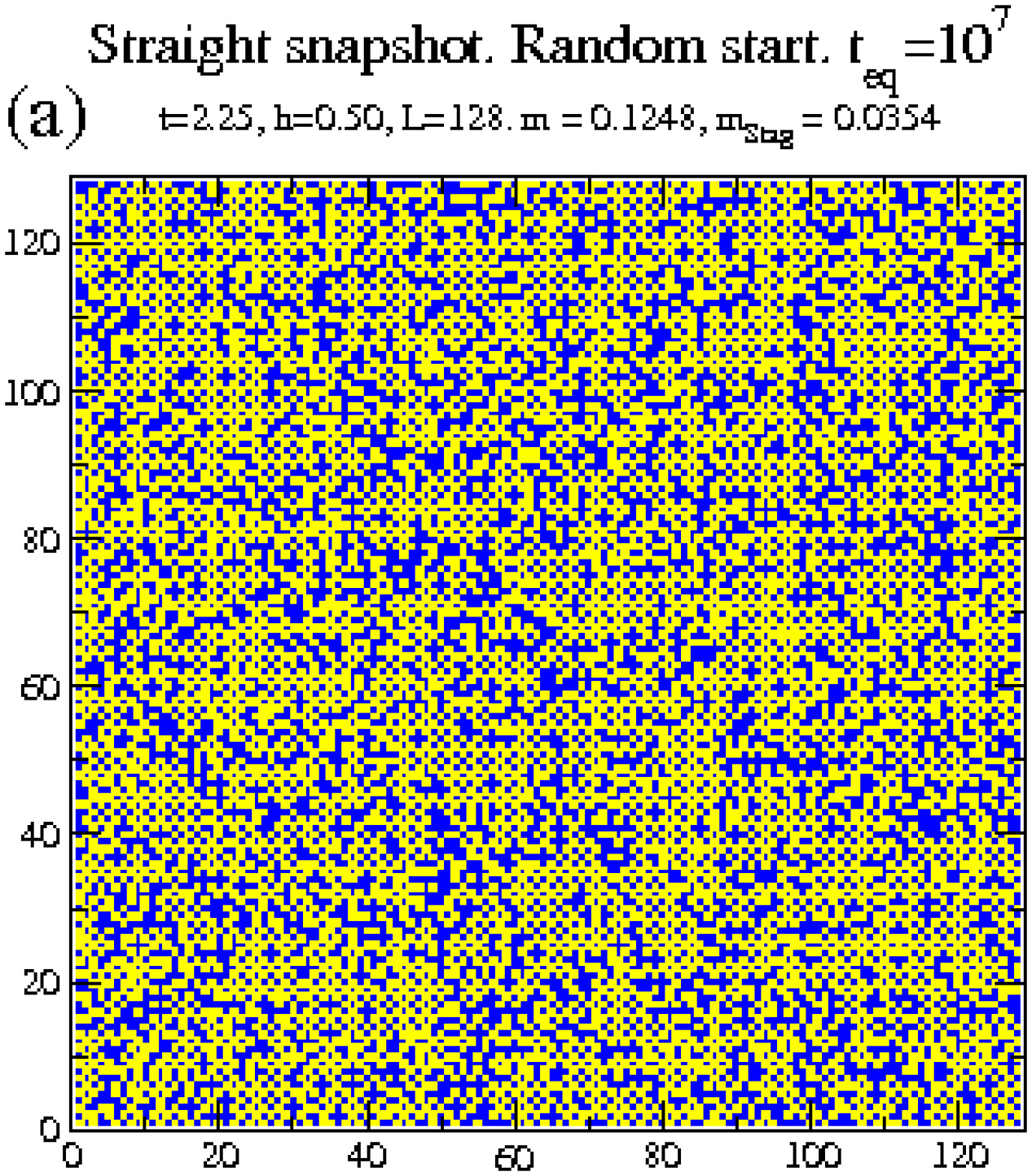} 
\hspace{0.6truecm}
\includegraphics[angle=0,width=.4\textwidth]{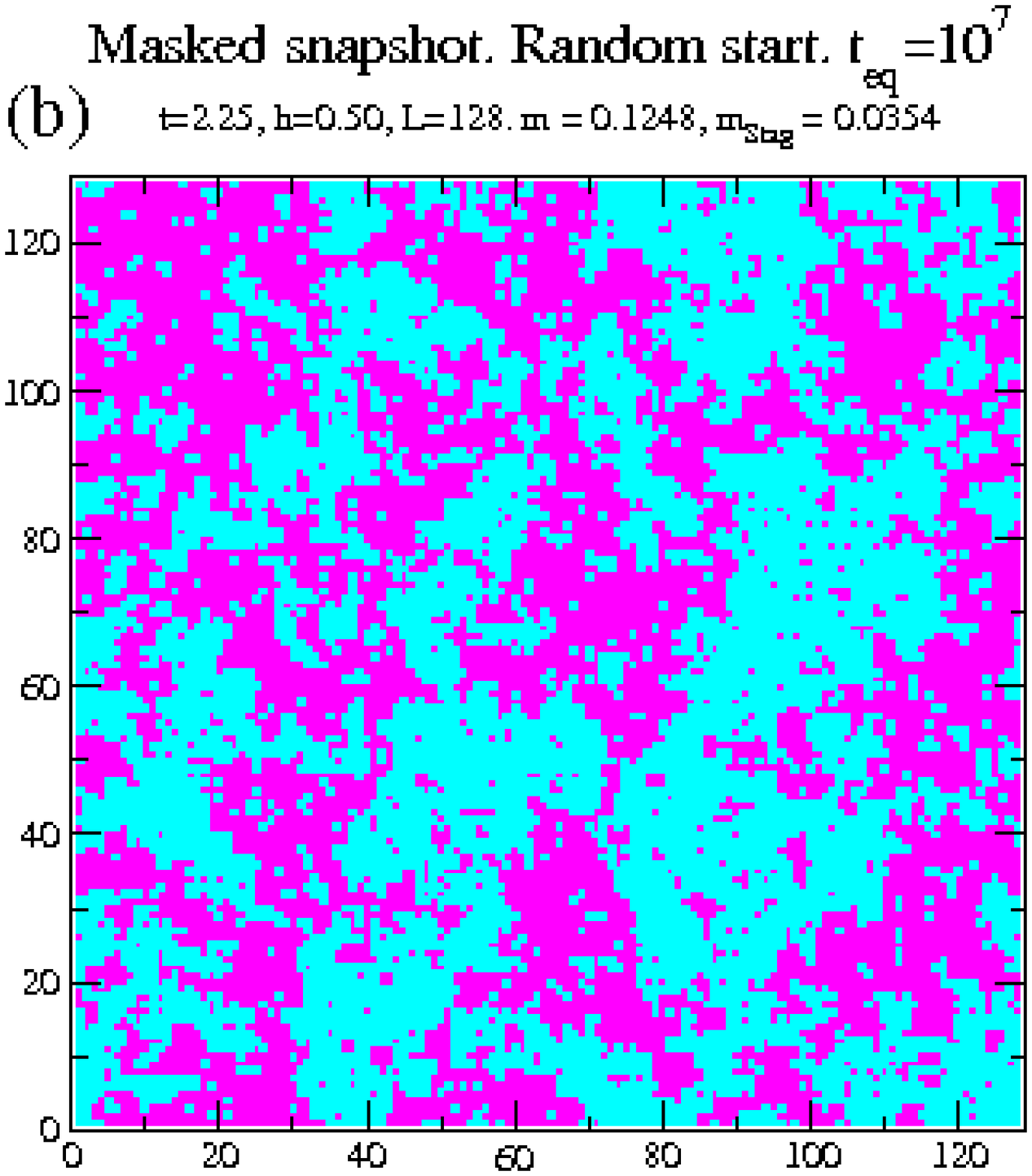} \\
\vspace*{1.0truecm}
\includegraphics[angle=0,width=.4\textwidth]{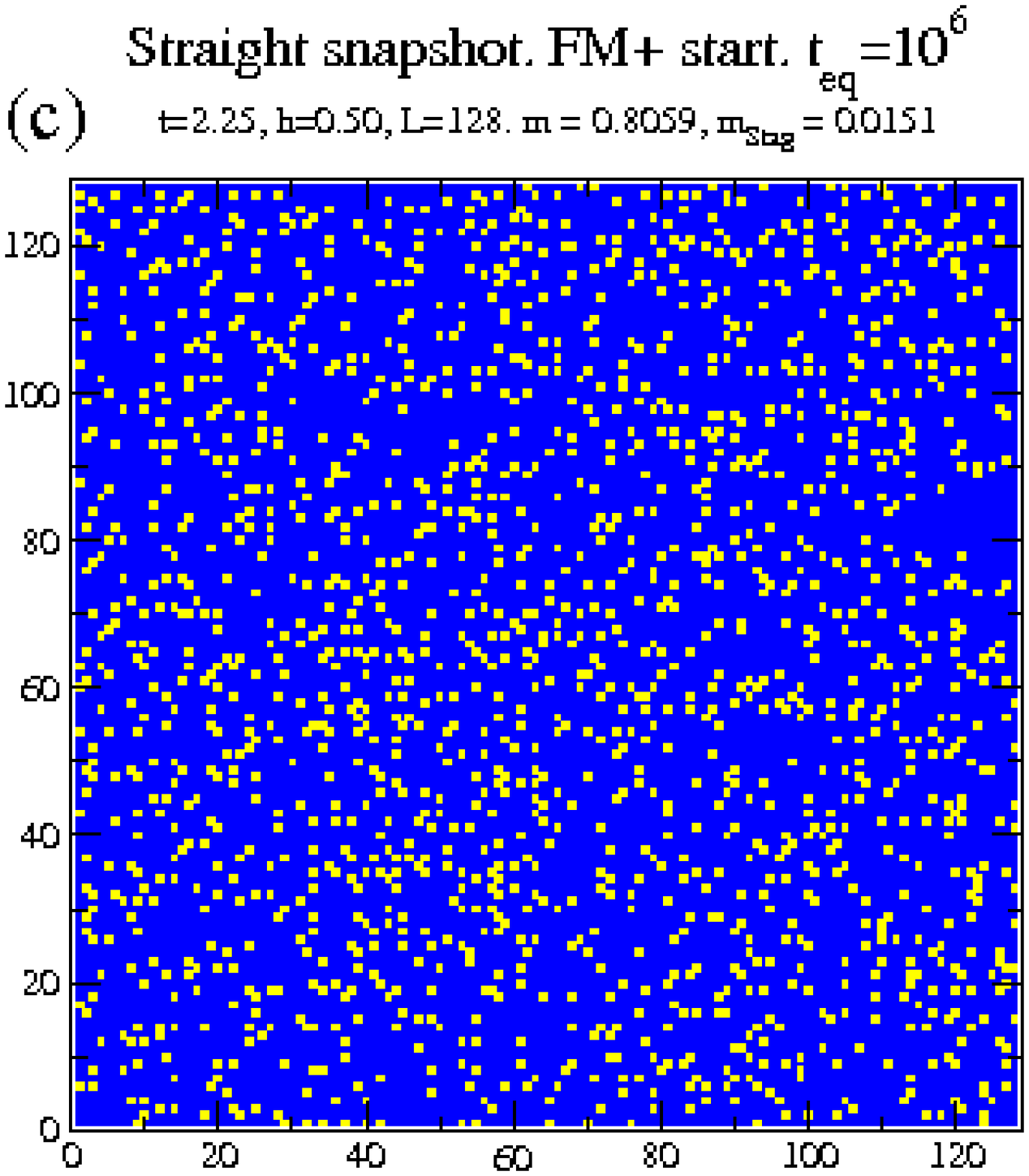} 
\hspace{0.6truecm}
\includegraphics[angle=0,width=.4\textwidth]{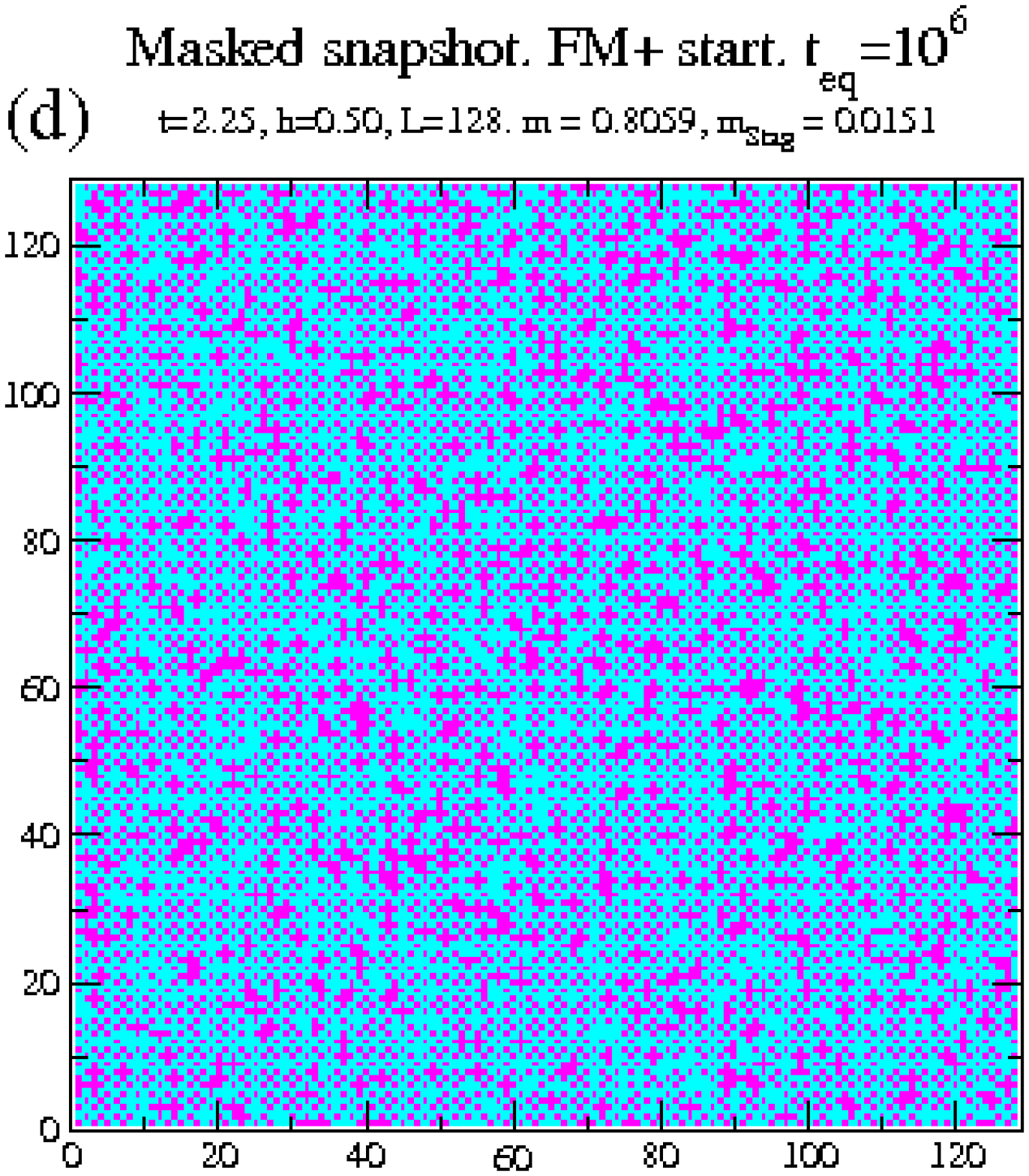}
\end{center}
\vspace*{1.5truecm}
\caption[]{
\baselineskip=0.15truecm
(Color online) 
Snapshots for $a=7$ at $t=2.25$ and $h=0.50$ in the horn region 
(marked by a diamond in Fig.\ \protect\ref{fig:PhD_A70}(b)). 
(a) and (b) show the equilibrium disordered phase 
with AFM fluctuations. The system was initiated in an uncorrelated, random configuration and equilibrated for 
$10^7$ MCSS before the image was recorded. 
(c) and (d) show the metastable FM+ phase. The system was initiated in the fully ordered FM+ configuration and 
``equilibrated" for $10^6$ MCSS. 
In the ``straight" images, (a) and (c), up and down spins are colored dark gray (blue online) 
and light gray (yellow online), respectively. 
The ``masked" images (b) and (d) emphasize AFM domains by coloring up spins on the A sublattice and down spins 
on the B sublattice dark gray (magenta online) and down on A, up on B light gray (cyan online). 
}
\label{fig:snap}
\end{figure}

The main part of Fig.~\ref{fig:PhD_A70}(b) shows a magnified image of the horn region. At this level of detail, two 
interesting phenomena become apparent. The first is that the Ising critical line (obtained from the crossings of fourth-order cumulants and linearly extrapolated to $1/L = 0$) 
continues beyond the critical endpoint where it meets the coexistence line, until it 
meets the spinodal line at $h \approx 0.705$. The triangular region limited by the coexistence line, the 
extended critical line, and the spinodal line is shown in further detail in Fig.~\ref{fig:PhD_A70}(c). 
Our interpretation of this structure is that the extended critical line represents a nonequilibrium, 
second-order, Ising phase transition between a metastable AFM phase at lower temperatures and a 
metastable disordered phase inside the triangle. 
The existence of a true phase transition between two metastable phases is possible because of the divergence of the 
metastable lifetimes in the thermodynamic limit. 
In this triangular region, FM+ has the lowest free energy and is therefore the stable phase. Conversely, in 
the rest of the horn region, the stable phase is the disordered phase, while FM+ is metastable. 
Snapshots of the stable and metastable phases in the latter region at the phase point marked by a diamond in 
Fig.~\ref{fig:PhD_A70}(b) are shown in Fig.~\ref{fig:snap}. (See methodological details in the caption of that figure.) 
In the triangular region in Fig.~\ref{fig:PhD_A70}(c), the stability of FM+ was confirmed by starting systems 
with $L$ up to 1024 with a completely random spin configuration and equilibrating at $t=2.15$ and $h=0.645$ 
 [marked by an up triangle in Fig.~\ref{fig:PhD_A70}(c)] for $10^7$ MCSS before measuring the order parameters. 
The metastability of the disordered phase 
in the same region was confirmed by scanning $h$ in the positive direction across the coexistence line until it 
decayed discontinuously to FM+ at the spinodal. At the point where the extended critical line meets the spinodal 
line, the interpretation of the latter changes from the limit of metastability of the AFM phase at lower $t$ to 
being the limit of metastability of the disordered phase at higher $t$. 
The line was determined by the same method in both temperature regions. 

The second phenomenon observed in Figs.~\ref{fig:PhD_A70}(b) and (c) is that, above $t \approx 2.20$, 
the coexistence line and the spinodal line for the disordered phase coincide within our numerical accuracy. 
The inset in Fig.~\ref{fig:PhD_A70}(b) demonstrates that the separation of the spinodals near the tip of each horn 
obeys  Eq.~(\ref{eq:hspin}).
At $h=0$ the critical temperature is again near the exact Ising value, $t_c(h=0) \approx 2.269$, unaffected by the 
long-range interaction. 

\subsubsection{Hysteresis loops}

\begin{figure}[ht]
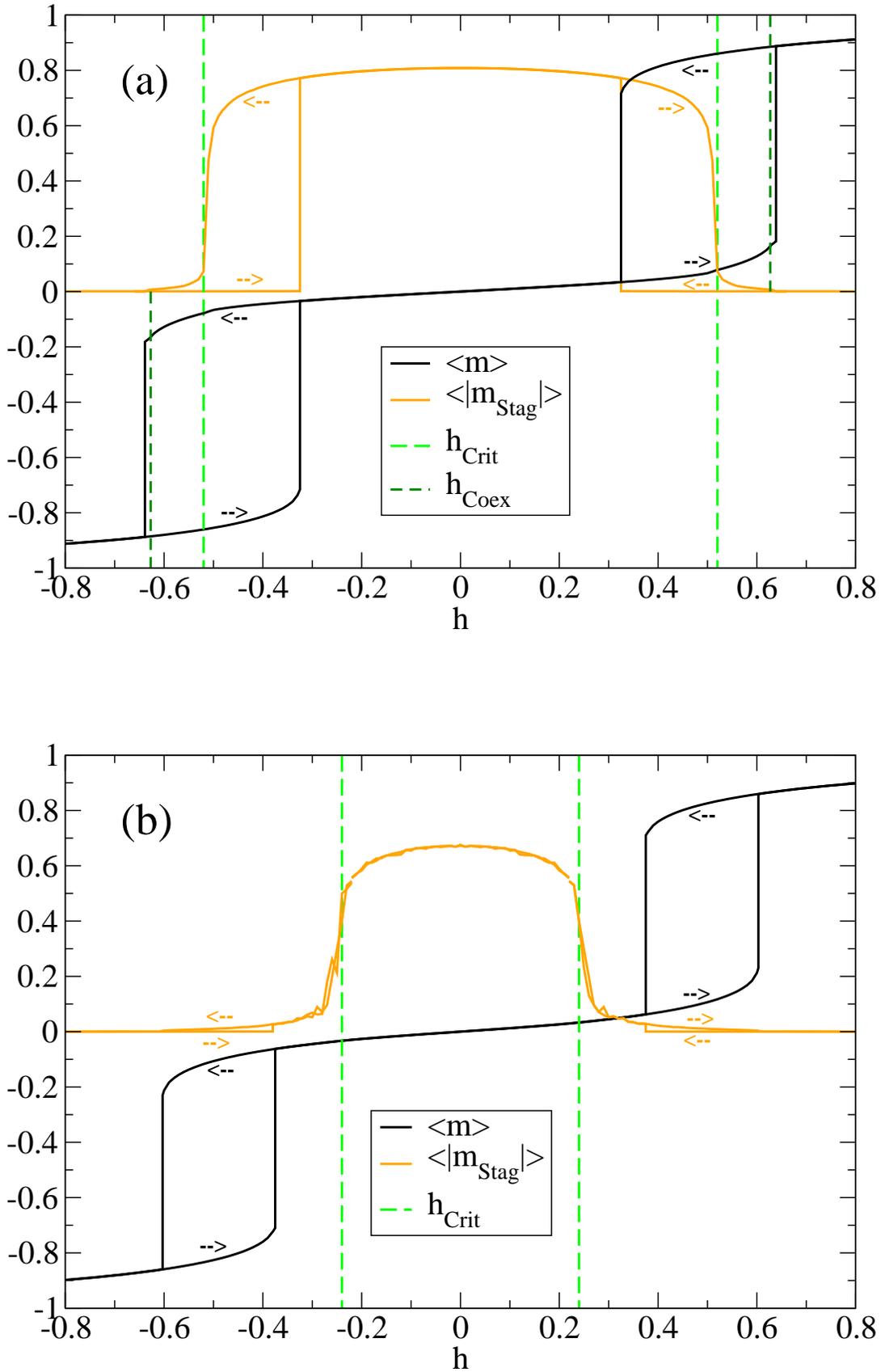

\begin{center}
\vspace*{-0.8truecm}
\includegraphics[angle=0,width=.8\textwidth]{Hyst_L512_1024_T218.eps} \\
\vspace*{1.8truecm}
\includegraphics[angle=0,width=.8\textwidth]{Hyst_L512_1024_T225.eps}
\end{center}
\vspace*{-0.5truecm}
\caption[]{
\baselineskip=0.15truecm
(Color online) 
Constant-temperature hysteresis loops for $a=7$. 
The system size was $L=512$, except near where the hysteresis path crosses the critical line 
(marked with light gray (green online) vertical, dashed lines) and in the disordered phase, where $L=1024$ was used. 
The nonzero values of $m_{\rm Stag}$ in the disordered phase are a finite-size effect. 
(a) 
At $t = 2.18$. The metastable FM+ phase decays discontinuously into the ordered AFM phase at 
the FM+ spinodal near $h = +0.34$. The loops are symmetric under reversal of $h$ and $m$. 
The dark, vertical, short-dashed lines mark the coexistence lines between the disordered and FM phases. At this 
temperature they lie very close to the disorder spinodals. 
(b) 
At $t = 2.25$. 
The metastable FM+ phase decays discontinuously into the disordered phase at 
the FM+ spinodal near $h = +0.39$. The loops are symmetric under reversal of $h$ and $m$. 
At this temperature, the disorder/FM coexistence lines and the disorder spinodals coincide within our 
numerical accuracy. 
See further discussion in the text. 
}
\label{fig:Hhyst}
\end{figure}

The phase diagram for $a=7$, shown in  Fig.~\ref{fig:PhD_A70}, suggests the existence of complex hysteresis loops. 
Constant-temperature hysteresis loops for two different temperatures in the horn region are shown in 
Fig.~\ref{fig:Hhyst}. In Fig.~\ref{fig:Hhyst}(a) we use $t = 2.18$, which lies between the temperature of the 
critical endpoint and the temperature at which the FM spinodal lines cross the critical line. Following the curves in the 
negative direction from $h=+0.8$, the system starts in the stable FM+ phase, which becomes metastable as the 
phase point crosses the coexistence line at $h \approx +0.63$. 
(Note that the disorder/FM coexistence lines are at only slightly weaker fields than 
the disorder spinodal lines at this temperature. See  Figs.~\ref{fig:PhD_A70}(b) and (c).) 
Crossing the FM+ spinodal line at $h \approx +0.33$, the system changes discontinuously to the 
equilibrium AFM phase. It remains in this phase until it crosses the critical line into the equilibrium disordered phase 
at $h \approx -0.52$. Finally there is a discontinuous change to the FM$-$ phase across the disorder 
spinodal at $h\approx -0.63$. 
The sequence is symmetric in $h$ as the field is reversed from $h=-0.8$ back to $+0.8$. 

Raising the temperature to $t=2.25$ (which lies between the temperature at which the FM spinodals cross the 
critical line and the zero-field critical temperature) in Fig.~\ref{fig:Hhyst}(b), the main difference is that the system changes 
discontinuously from the FM+ phase to the equilibrium disordered phase at $h \approx +0.38$, only passing into the 
equilibrium AFM phase as it crosses the critical line at $h\approx +0.24$. At $h\approx -0.24$ it again crosses the critical 
line into the disordered phase, which it leaves through a discontinuous jump as it crosses out of the 
horn region at the negative disorder spinodal at $h\approx -0.60$. The path is again symmetric during the field reversal. 
At both temperatures the nonzero values of the staggered magnetization in the disordered equilibrium phase are a 
finite-size effect. 

\begin{figure}[ht]
\begin{center}
\vspace*{-0.8truecm}
\includegraphics[angle=0,width=.8\textwidth]{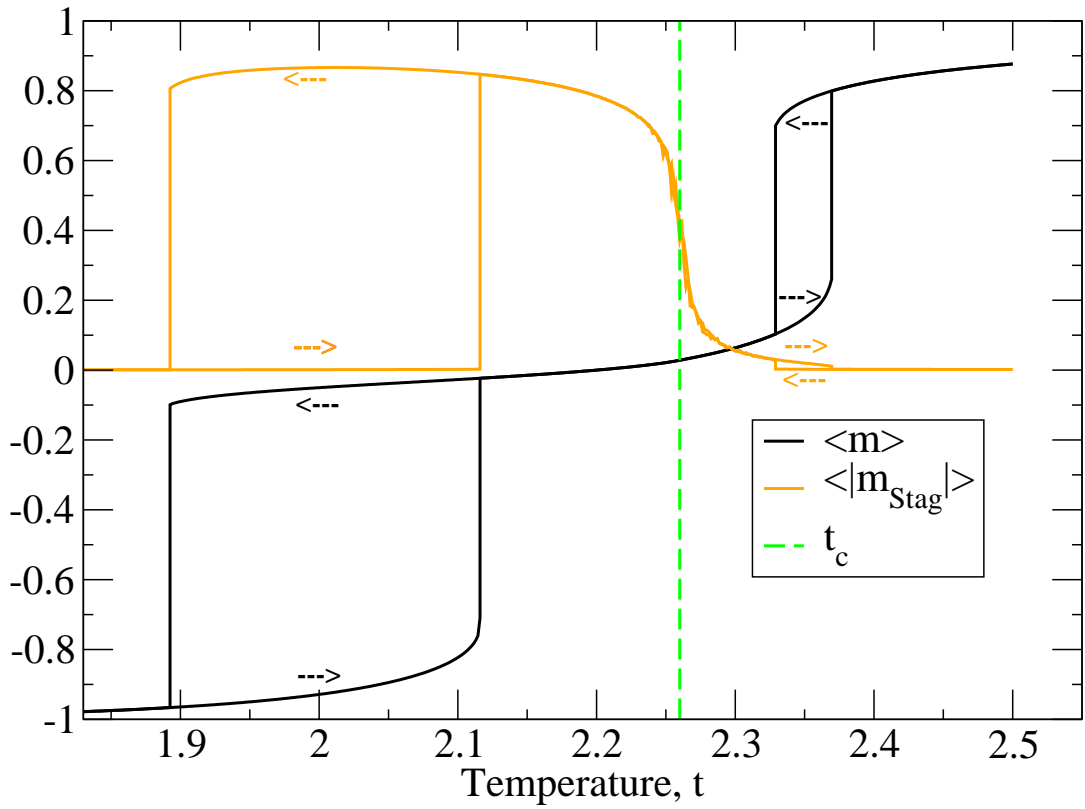}
\end{center}
\vspace*{1.5truecm}
\caption[]{
\baselineskip=0.15truecm
(Color online) 
Asymmetric thermal hysteresis loops for $a=7$ 
along the path marked by a diagonal line segment in Fig.~\protect\ref{fig:PhD_A70}(a). 
The vertical, dashed line marks the temperature where the path crosses the line of AFM Ising critical points. 
Simulated with $L=256$.  The nonzero values of $m_{\rm Stag}$ in the disordered phase are a finite-size effect. 
This pattern of transitions and hysteresis loops closely 
resembles recent experimental results for thermal two-step transitions with 
hysteresis in several different SC materials.\protect\cite{BONN08,PILL12,BURO10,LIN12,KLEI14,HARD15} 
See further discussion in the text. 
}
\label{fig:Thyst}
\end{figure}

When using the model studied here to represent phase transitions in SC materials, the magnetic field is replaced by the 
temperature-dependent effective field of Eq.~(\ref{eq:heff}). A path for temperature driven hysteresis within this 
interpretation of the model is represented by the diagonal line segment in Fig.~\ref{fig:PhD_A70}(a) 
(degeneracy ratio $\ln g = 20/3$ and energy difference $D = 18|J|$). The corresponding phase transitions and 
hysteresis loops are shown in 
Fig.~\ref{fig:Thyst}. The loops are asymmetric. The narrow loop above the critical temperature corresponds
to passage across the positive-$h$ horn, while the wider loop below the critical 
temperature lies between the negative AFM spinodal 
and the FM$-$ spinodal.  The nonzero values of $m_{\rm Stag}$ in the disordered phase region are again a finite-size 
effect. 

We note that the pattern of transitions and hysteresis loops shown in Fig.~\ref{fig:Thyst} closely resembles recent experimental results for thermal two-step transitions with 
hysteresis in several different SC materials.\cite{BONN08,PILL12,BURO10,LIN12,KLEI14,HARD15} 
Some of these experiments are also discussed in two recent reviews of this rapidly developing field.\cite{SHAT15,BROO15}

\subsection{$a = 10$}
\label{sec:4C}

\begin{figure}[ht]
\begin{center}
\vspace*{-0.8truecm}
\includegraphics[angle=0,width=.8\textwidth]{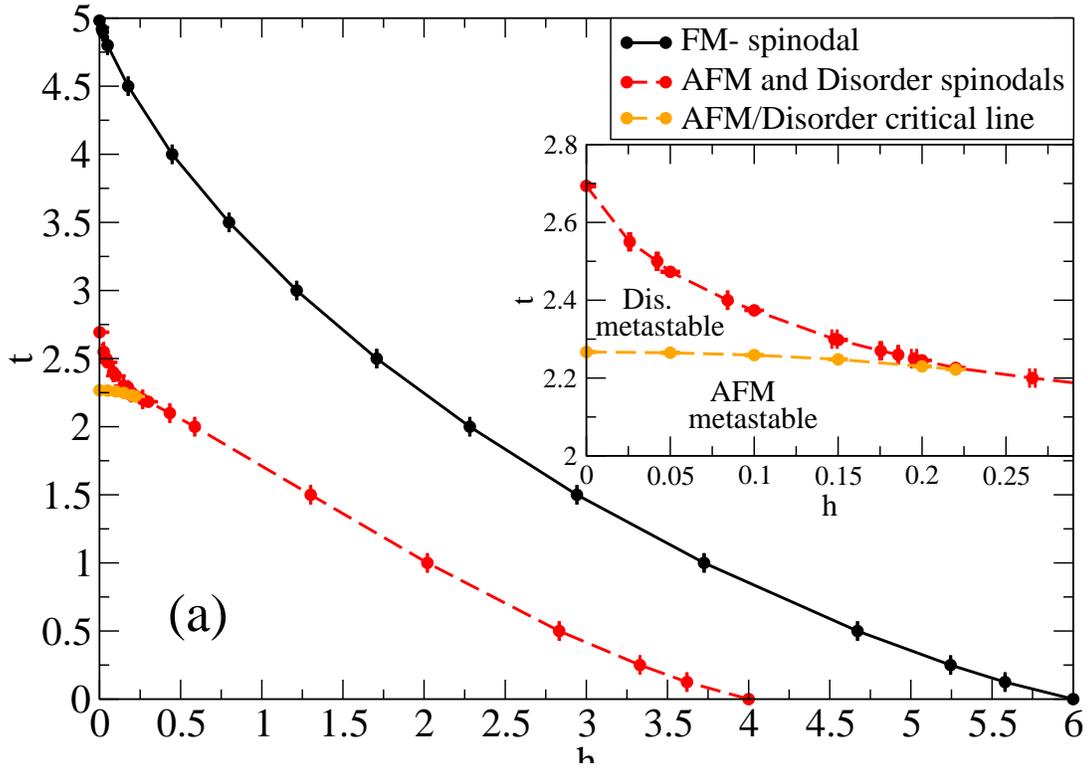} \\
\vspace*{1.5truecm}
\includegraphics[angle=0,width=.8\textwidth]{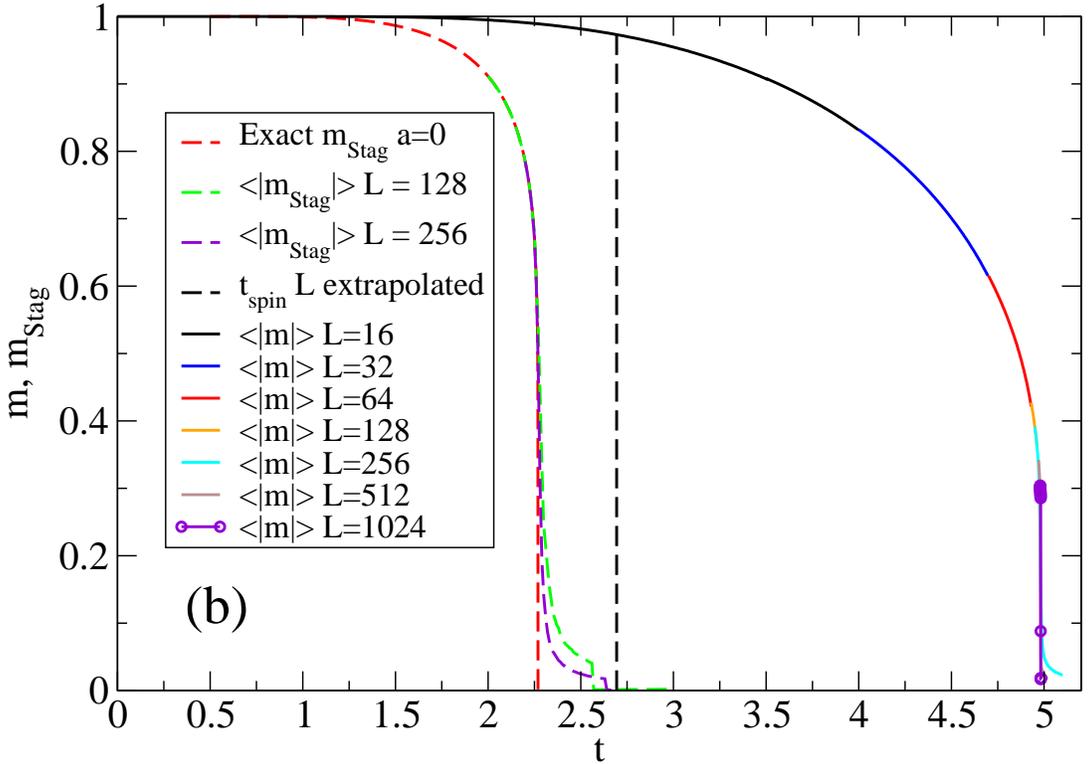}
\end{center}
\vspace*{-0.5truecm}
\caption[]{
\baselineskip=0.15truecm
(Color online) 
(a)
$h,t$ phase diagram for $a=10$. 
The FM+ phase is stable everywhere in this figure, and FM$-$ is metastable everywhere below the black spinodal curve. 
The AFM phase is metastable below the medium gray (red online) spinodal and light gray (orange online) critical curves, and the disordered phase is metastable inside the triangle between these curves. 
The inset shows an enlarged view of the region around the metastable disordered phase. 
The phase diagram is symmetric under simultaneous sign change of $h$ and exchange of the FM+ and FM$-$ phases. 
(b)
Stable FM order parameter $m$ (solid curves) and metastable AFM order parameter $m_{\rm Stag}$ (dashed curves), 
shown vs $t$ at $h=0$. The FM order parameter is shown for a composite of system sizes, $L = 16$, ... 1024, identified 
online by different colors. 
The data for $L = 512$ and 1024 indicate that the transition is weakly first-order. 
The AFM order parameter is shown as the exact, Onsager order parameter and MC 
simulations for $L = 128$ and 256. The metastable disordered phase lies between the Ising critical temperature and 
the $L$-extrapolated spinodal temperature marked by the vertical dashed line. 
See further discussion in the text. 
}
\label{fig:PhD_A100}
\end{figure}

For $a > 8$ there is no stable AFM phase. The $h,t$ phase diagram with $a=10$ is shown for $h \ge 0$ 
in Fig.~\ref{fig:PhD_A100}(a). The phase diagram is symmetric around $h=0$.
The black curve with data points in the main figure shows the FM$-$ spinodal. For weak fields and temperatures below 
the Ising critical temperature ($t_{c}(h=0) \approx 2.269$), there also exists a metastable AFM phase. 
It is separated from the stable 
FM phases by sharp mean-field spinodals. At higher temperatures, this AFM phase undergoes a second-order transition to a metastable disordered phase.  These metastable phases are separated by a line of critical points in the Ising 
universality class. (Located and identified by the fourth-order cumulant method as above.) 
At $h=0$, the disordered phase is metastable between the zero-field Ising critical temperature 
and $t \approx 2.68$.  
Like the metastable AFM phase, it is separated from the equilibrium FM phases by sharp mean-field spinodals. 
A magnified view of the region containing the metastable disordered phase is shown in the inset in 
Fig.~\ref{fig:PhD_A100}(a).
The spinodals and the metastable critical line shown in Fig.~\ref{fig:PhD_A100}(a) were obtained from finite-size 
scaling extrapolations of MC data up to $L = 1024$. These features do not appear in the simple mean-field 
approximation shown in  Fig.~\ref{fig:MFA_A100}(a). 
Stable and metastable order parameters at $h=0$ are shown in Fig.~\ref{fig:PhD_A100}(b), 
corresponding to the mean-field results shown in Fig.~\ref{fig:MFA_A100}(b). 
The MC simulations for the metastable order parameter are seen to be in excellent agreement with the 
Onsager-Yang exact 
order parameter for a square-lattice Ising model in zero field,\cite{PATH11} 
$m_{\rm Stag}(t,0) = \left\{ 1 - \left[ \sinh (2/t)^{-4} \right] \right\}^{1/8}$. 
The metastable disordered phase is characterized by values of the simulated $m_{\rm Stag}$ that decrease 
linearly with $L$ and go discontinuously to zero at a sharp spinodal temperature. 
This phase lies between the Ising critical temperature and the spinodal temperature, whose $L$-extrapolated value is  
marked by a vertical, dashed line. 
If the system is heated in the metastable phases, this is the temperature at which the stable order parameter $m$ 
will jump discontinuously from near zero to its equilibrium value. 
The observation of a critical line separating the metastable AFM and disordered phases in this large-$a$ regime 
supports our interpretation of the critical line in the 
small part of the horn region for $a=7$, shown in Fig.~\ref{fig:PhD_A70}(c) as a transition line separating two 
metastable phases. 

The phase transition of the stable FM phase at $t \approx 4.98$ involves a small discontinuity (seen only for $L = 1024$) 
and negative values of the Binder cumulants above the transition temperature (seen for $L = 1024$ and 512, 
not shown). These features suggest that this transition is weakly first-order. 
Exploratory simulations for $a = 8.5$ and $a = 20$ at $h=0$ indicate a strongly first-order transition in the former case, 
and a continuous transition in the mean-field universality class in the latter. Further investigation of the strong 
long-range interaction regime of $a \ge 8$ is left for future study.

\section{Conclusions}
\label{sec:CONC}

In this paper we present a detailed investigation of the phase diagrams of a simplified model of an SC 
material with a two-step transition as a square-lattice Ising model with AFM nearest-neighbor interactions and FM long-range 
interactions of the Husimi-Temperley (equivalent-neighbor) kind. An AFM equilibrium phase for weak applied fields 
is replaced by field-induced FM phases at stronger fields. These phases are separated by coexistence lines surrounded 
by sharp spinodal lines representing limits of metastability. 

In a range of intermediate-strength 
long-range interactions, we find significant differences between the phase diagrams of this model, 
calculated by importance-sampling MC simulations, and those of a model 
in which fluctuations have been neglected by replacing the nearest-neighbor interactions by a two-sublattice 
mean-field approximation. The difference consists in the replacement of each tricritical point in the mean-field 
model with a pair consisting of a critical endpoint and a mean-field critical point in the $h,t$ phase diagram, 
surrounded by horns representing metastable phases. 
For even stronger long-range interactions, the AFM equilibrium phase disappears. 
However, metastable AFM and disordered phases can still be observed. 
These complex phase diagrams give rise to hysteresis loops reminiscent of two-step transitions observed in several SC 
materials.\cite{KOPP82,ZELE85,PETR87,BOUS92,JAKO92,REAL92,BOIN94,BOLV96,CHER04,BONN08,PILL12,BURO10,LIN12,KLEI14,HARD15,CHER03,HUBY04,SHAT15,BROO15} 

We find it likely that the horn type phase diagrams and related two-step hysteresis loops 
revealed in our model by MC simulations may be observed in 
more realistic models of SC materials with elastic interactions,\cite{NISH13} and also in future experiments.  
An investigation into the former possibility is in progress.\cite{NISH15} 
Other interesting avenues of further research include cluster mean-field approximations for the short-range AFM 
interactions \cite{ENDNOTE} and calculation of free-energy surfaces in the $m_{\rm A}, m_{\rm B}$ plane by 
Wang-Landau MC simulations.\cite{BROW15}

\section*{Acknowledgments}

P.A.R.\ and C.O.\ gratefully acknowledge hospitality by the Department of Physics, University of Tokyo. 
P.A.R.\ thanks H.R.D.\ Barclay for useful suggestions.
Work at Florida State University was supported in part by U.S.\ NSF Grant No. DMR-1104829. 
G.B.\ is supported by Oak Ridge National Laboratory, which is managed by UT-Battelle, LLC. 
The work was also supported by 
Grants-in-Aid for Scientific Research C (Nos.\ 26400324 and 25400391) from MEXT of Japan, 
and the Elements Strategy Initiative Center for Magnetic Materials under the outsourcing project of MEXT. 
The numerical calculations were supported by the supercomputer center of ISSP of The University of Tokyo 
and the Florida State University Research Computing Center.



\clearpage

\begin{thebibliography}{51}
\bibitem[{\citenamefont{Omand and Miyashita}()}]{ENDNOTE}
  \bibinfo{note}{Better agreement between mean-field and MC
  results can be obtained by including some fluctuation effects through cluster
  mean-field approximations with clusters of at least $6 \times 6$ spins.\cite{OMAN16}
  However, this approach will not change the universality class of critical lines.}
\end{thebibliography}

\end{document}